\documentclass[12pt,a4paper]{article}

\usepackage{float} 
\usepackage{amsmath}
\usepackage{amsfonts}
\usepackage{amssymb}
\usepackage{graphicx}
\usepackage{color}
\usepackage{booktabs}
\usepackage{inputenc}
\usepackage[T1]{fontenc}
\usepackage{mathrsfs}
\usepackage{enumerate}
\usepackage{xcolor,url}
\usepackage{cancel,dsfont} 
\usepackage{multirow}
\usepackage{soul}
\usepackage{bm}
\usepackage{braket}
\usepackage{subcaption}
\usepackage{listings}
\usepackage[para]{footmisc} 
\usepackage{url}
\usepackage{placeins}

\usepackage[normalem]{ulem}
\usepackage{nicematrix}

\definecolor{mypink1}{rgb}{0.858, 0.188, 0.478}

\newcommand{\eea}{\end{eqnarray}}
\newcommand{\bea}{\begin{eqnarray}}
\newcommand{\be}{\begin{equation}}

\usepackage[colorlinks,bookmarks]{hyperref}
\definecolor{linkblue}{rgb}{0,0,0.8}
\definecolor{linkgreen}{rgb}{0,0.5,0}

\hypersetup{pdfpagemode=UseNone, pdfstartview=FitH, linkcolor=linkblue,
            citecolor=linkgreen, urlcolor=linkblue}

\begin{document}

\begin{center}

{\Large \bf Probing Dark Relativistic Species and Their Interactions with Dark Matter through CMB and 21cm surveys}  \\[0.7cm]

{\large  Hugo Plombat,${}^{1}$ Th\'eo Simon,${}^{1}$  Jordan Flitter,${}^{2}$ Vivian Poulin${}^{1}$\\[0.7cm]}

\end{center}

\begin{center}

\vspace{.0cm}

{\normalsize { \sl $^{1}$ Laboratoire Univers \& Particules de Montpellier (LUPM), CNRS \& Universit\'e de Montpellier (UMR-5299), Place Eug\`ene Bataillon, F-34095 Montpellier Cedex 05, France}}\\
{\normalsize { \sl $^{2}$ Physics Department, Ben-Gurion University of the Negev, Beer-Sheva 84105, Israel}}
\vspace{.3cm}

\end{center}

\hrule \vspace{0.3cm}
{\small  \noindent \textbf{Abstract} \vspace{.1in}
\hrule
\vspace{0.3cm}
We investigate the sensitivity of the 21cm power spectrum from cosmic dawn and the epoch of reionization to models of free-streaming dark radiation (parameterized through $N_{\rm eff}$) and interacting dark radiation-dark matter models (DM-DR). The latter models have gained attention for their potential in addressing recent cosmological tensions and structure formation challenges. 
We perform a Fisher matrix analysis under different assumptions regarding the astrophysical modeling, and forecast the sensitivity of HERA observations, combined with CMB data from Planck and the Simons Observatory (SO), to $N_{\rm eff}$ and  DM-DR interaction modeled using the ETHOS framework assuming a constant scattering rate between the two components.
Most importantly, we find that 21cm observations can improve the sensitivity to the DM-DR interaction rate by up to four order of magnitude compared to Planck and SO. Conversely, in the limit of low interaction rate (which asymptotically matches $N_{\rm eff}$), CMB data dominates the constraining power, but the inclusion of HERA data can provide a $\sim 20\%$ improvement in sensitivity over CMB data alone. 
Moreover, we find that HERA observations will be able to probe a region of the DM-DR interaction parameter space which is promising to explain the weak lensing amplitude `$S_8$' tension. Our results demonstrate the complementarity of 21cm and CMB data in exploring dark sector interactions.  
}

\vspace{0.3cm}
\newpage

\tableofcontents

\section{Introduction\label{sec:intro}}
The existence of cold dark matter (CDM), presenting solely gravitational interactions, is strongly supported by a wide-variety of observations, from the galactic and galaxy cluster scale~\cite{Zwicky:1933gu,rubin70,randall08} to the cosmological scale, and in particular by the Cosmic Microwave Background (CMB)~\cite{Planck:2018vyg}, and large-scale structures surveys. 
Yet, the search for the fundamental nature of dark matter has been ongoing for several decades, and despite strong constraints (and some hints) from ever-more precise observations, remains largely unknown. However, small-scale observations come up against various challenges when comparing  with the results of N-body simulations, such as the core-cusp problem~\cite{deBlok:1996jib,Walker:2011zu,Donato:2009ab} (the observed density profile of halos looks core-like at the center, while simulations prefer cuspy profiles), diversity problems~\cite{Oman:2015xda} (the observed velocity profiles in dwarf galaxies present more diversity than predicted by simulations), and  the ``too-big-to-fail'' problem~\cite{Boylan-Kolchin:2011qkt,Papastergis:2014aba} (the most massive predicted sub-halo satellites of Milky Way remain unobserved while they should host luminous galactic content). The integration of baryonic physics could partially solve several problems~\cite{Pontzen:2011ty,Garrison-Kimmel:2017zes}, but a consensus has yet to emerge in the community, motivating the exploration of alternative models for dark matter potentially capable of overcoming the issue encountered by CDM. 
In addition, recent weak-lensing experiments~\cite{Heymans:2020gsg} have indicated a smaller amplitude of fluctuation on scales of $\sim 1$ Mpc, characterized by the parameter $S_8\equiv \sigma_8\sqrt{\Omega_m/0.3}$, than predicted in the $\Lambda$CDM model fit to Planck CMB data~\cite{Planck:2018vyg}, at the $2-3\sigma$ level. Similarly to the small-scale problem of DM, it has been suggested that improvements in the modelling may resolve the tension \cite{Arico:2023ocu}, but this remains largely debated in the literature \cite{Amon:2022azi,Salcido:2024qrt}.\\

In recent years, models proposing a coupling between DM and a light thermal bath of dark radiation (DR) have gained in interest in light of these issues of the CDM paradigm. These models, initially proposed by refs.~\cite{Boehm_2001,Boehm_2004}, have subsequently been widely explored~\cite{Ackerman:2008kmp,Chu:2014lja,Buen-Abad:2015ova}, and constrained using CMB data~\cite{Lesgourgues:2015wza,Archidiacono:2017slj} or large-scale structure data~\cite{Rubira:2022xhb,Mazoun:2023kid,Mazoun:2024uad}. DM-DR interaction can significantly affect the formation of structures, by causing suppressions and/or oscillations in the matter power spectrum at small scales. The effective theory of structure formation (ETHOS)~\cite{Cyr-Racine:2013fsa,Bohr:2020yoe} has been formulated to develop a mapping between the DM particle physics models and their physical effects on the matter power spectrum, via effective generic parametrisations.
The presence of additional relativistic degrees of freedom, quantified by $N_{\rm eff}$ (or $N_{\rm idr}$ when the extra relativistic species is modelled as interacting), can further alter the shape of the CMB and matter power spectra, providing another handle on those models.
Interestingly, it has been shown that the $S_8$ tension may also be explained by models of DM-DR interaction~\cite{Archidiacono:2019wdp}. In addition, most of these models lead to very strong imprints at small scales, and can be particularly well measured (or constrained) by late time probes sensitive to the scales affected by DM-DR interaction. This includes for instance the well-known Lyman-$\alpha$ forest flux power spectrum at $z\lesssim 5$, whose data have already been used to constrain this class of models~\cite{Archidiacono:2019wdp}. \\

Alternatively, the 21cm signal from Cosmic Dawn (CD) is also a very sensitive probe of dark matters properties, as studied in, \textit{e.g.}, refs.~\cite{Lopez-Honorez:2016sur,Lopez-Honorez:2018ipk,Escudero:2018thh,Munoz:2018jwq,Munoz:2018pzp,Sun:2023acy,Facchinetti:2023slb,Qin:2023kkk,Flitter:2022pzf,Giri:2022nxq}.
 Processes leading to an alteration of the matter power spectrum at small scales  can affect the formation of first stars, whose emitted Lyman-$\alpha$ photons  play a considerable role in the 21cm signal via Wouthuysen-Field effect~\cite{wouthuysen52,field58}, but also shift the entire history of CD and the Epoch of Reionization (EoR).
These effects could hopefully be measured with the HERA interferometer~\cite{DeBoer:2016tnn}, which is dedicated to the observations of fluctuations in the 21cm signal during CD ($z\sim 10 - 30$) and EoR ($z \sim 5 - 10$), providing the current most constraining upper bound on the 21cm power spectrum~\cite{HERA:2022wmy}. \\

In this article, we present forecasts, using Fisher matrix analysis,  on the sensitivity of 540 days HERA observation on a model of DM-DR interaction.  As a preliminary study and in order to disentangle the effect of the additional radiation from that of the interaction, we first study the sensitivity to free-streaming $N_{\rm eff}$ of the 21cm signal. This study is complementary to the one carried by ref.~\cite{Lee:2023uxu}, which proposes forecasts on  the combined sensitivity of the Square Kilometer Array, CMB-Stage4, and DESI Baryon Acoustic Oscillations observations to $N_{\rm eff}$, investigating its degeneracy with the primordial helium fraction. We then study the case of interacting DM-DR, using the ETHOS parametrisation in the case of a constant scattering rate of DR over DM, that is favored by recent weak lensing data, and to which the 21cm signal is expected to be more sensitive. \\

We carry out our study using \texttt{21cmFirstCLASS}~\cite{Flitter:2023mjj,Flitter:2023rzv}, a version of \texttt{21cmFAST}~\cite{Mesinger:2010ne,Murray:2020trn} interfaced with the Boltzmann code \texttt{CLASS}~\cite{Lesgourgues:2011re}, providing us the initial conditions of the simulation. We perform our analysis on cosmological and astrophysical parameters, and estimate their degeneracies.  Lastly, we combine the 21cm mock data with Planck~\cite{Planck:2018vyg} and forthcoming Simons Observatory (SO)~\cite{SimonsObservatory:2018koc} CMB observations, in order to establish the complementarity of these two probes in constraining DM-DR interactions. Our analysis is complementary to the recent ref.~\cite{Verwohlt:2024efh}, that explores different models of DM-DR within the ETHOS framework and make use of different astrophysical models as implemented in \texttt{Zeus21}~\cite{Munoz:2023kkg}. \\

This paper is organized as follows. In section~\ref{sec:method} we briefly review how the 21cm signal is modelled, and present the Fisher matrix formalism as well as the data employed in our analysis. Section~\ref{sec:neff_section}  is dedicated to the forecast of the sensitivity to the effective number of relativistic species modelled with $N_{\rm eff}$. We then present in section~\ref{sec:idmdr_section} our main forecasts on a DM-DR interaction model that appears promising to explain the $S_8$ tension, and conclude in section~\ref{sec:conclusion}.

\section{Method \& Data}
\label{sec:method}
\subsection{Theoretical framework}

The 21cm hyperfine hydrogen line signal is usually quantified by the brightness temperature, \textit{i.e.} the blackbody temperature associated with the 21cm specific intensity.  The observable of interest is actually the  contrast in temperature between a patch of intergalactic medium (IGM) and  the radio background, which we take here to be the CMB. 
The differential brightness temperature can be expressed as~\cite{Pritchard:2011xb},
\begin{align}
    \delta T _b  &= \frac{T_s -T_{CMB}}{1+z}\left(1 - e^{-\tau_{\nu}}\right)\\
    & \approx 27 x_{HI} (1+\delta_b) \left(\frac{\Omega_b h^2}{0.023} \right) \sqrt{\frac{0.15}{\Omega_m h^2} \frac{1+z}{10}} \left( 1+ \frac{1}{H}\frac{d v_r}{dr} \right) ^{-1} \left( 1 - \frac{T_{\gamma}}{T_s}\right) \, ,
    \label{eq:tb}
\end{align}
where $T_{\gamma} \propto (1+z)$ is the CMB temperature and  $T_s$ is the spin temperature of the neutral hydrogen gas, which measures the population ratio between the hyperfine excited and ground states of neutral hydrogen. In addition, $\tau_{\nu}$  is the 21cm optical depth,
$x_{HI}$  is the hydrogen neutral fraction, and $\delta_b$ is the relative density perturbation in the relative baryon density. The differential brightness temperature is expressed here  so as to make explicit part of its dependence on cosmological parameters, with $\Omega_b$  and $\Omega_m$, the baryon and matter energy density fraction today, and $h$, the Hubble parameter in units of 100 km/s/Mpc. Finally $H(z)$ is the Hubble rate and $dv_r/dr$ is the gradient of the line of sight peculiar velocity, describing a relative motion of the gas with respect to the Hubble flow. \\

To model the 21cm-signal, we use the version \texttt{21cmFirstCLASS}~\cite{Flitter:2023mjj,Flitter:2023rzv}\footnote{\url{https://github.com/jordanflitter/21cmFirstCLASS} This work is carried out on a previous version of the code publicly available in the directory, with minor differences that will not affect our results.} of the \texttt{21cmFAST} code~\cite{Mesinger:2010ne,Murray:2020trn}. 
This version is interfaced with the Boltzmann code \texttt{CLASS}~\cite{Lesgourgues:2011re}, which allows one to generate initial conditions for \texttt{21cmFAST} which are consistent with the chosen cosmological model.  
The information employed in our analysis are the fluctuations in the 21cm signal. Writing the brightness temperature fluctuations as $\delta_{21}(\bm{x}) = (\delta T_b(\bm{x}) - \delta\bar{T_b})/\delta\bar{T_b}$, $\delta\bar{T_b}$ being the  sky averaged brightness temperature (also referred to as global signal), the 21-cm power spectrum is defined as the angle-averaged sum of the Fourier transform of $\delta_{21}(\bm{x})$:
\begin{equation}
    \langle \delta_{21} (\bm{k_1}) \delta_{21}^*(\bm{k_2}) \rangle = (2\pi)^3 \delta_D(\bm{k_1- k_2})P_{21}(\bm{k}) \, ,
\end{equation}
where $\delta_D$ is the Dirac delta function, $\langle \rangle$ the ensemble average, and $*$ the complex conjugate. 
In particular, in the following, we work with the reduced power spectrum (in unit of mK$^2$), defined as
\begin{equation}
    \Delta^2_{21}(k) = \frac{k^3P_{21}(k)}{2\pi^2} \, .
\end{equation}
The 21-cm power spectrum is sensitive to the variations of all the components in the brightness temperature: the underlying baryonic matter density field, the ionization fraction, the Ly-$\alpha$ coupling coefficient, the neutral hydrogen gas temperature, and the line-of-sight peculiar velocity gradient.\\

\texttt{21cmFAST} simulates the astrophysics evolution, such as the Lyman$-\alpha$ photons emissions,  X-ray heating of the IGM, or inhomogeneous reionization, employing a parametric model. We are interested in  the 21cm signal during all the stages from the CD to reionization, and base our model on the EOS2021 simulation parametrization~\cite{Munoz:2021psm}, which represents the current state-of-knowledge of the evolution of cosmic radiation fields during the first billion years. This model includes two populations of galaxies. Atomic cooling galaxies (ACGs) are populated mainly by Pop~II-type stars,  and molecular cooling galaxies (MCGs), hosted by minihalos (of virial mass $M_{\rm vir} \lesssim 10^8 M_\odot$), feature Pop~III-type stars, the first generation of stars, characterised by very low metallicity. These galaxies are expected to form earlier than the ACGs, due to the excitation of the roto-vibrational levels of the $\rm H_2$ molecule, which induces cooling of the gas within the halos (see \textit{e.g } refs.~\cite{Tegmark:1996yt,Haiman:2006si,Trenti:2009cj,Glover:2012gx,Bromm:2013iya,mebane13}). \\

In the end, in the the following analysis, we consider 10 free astrophysical parameters:
\begin{equation}
\begin{split}
    \theta_{\rm astro} = &\{\log_{10}(f_{*,10}^{II}), \alpha_*^{II}, \log_{10}(f_{\rm esc,10}^{II}) , \log_{10}(L_{X}^{II}), \\
    &\log_{10}(f_{*,7}^{III}), \alpha_*^{III}, \log_{10}(f_{\rm esc,7}^{III}) , \log_{10}(L_{X}^{III}), \alpha_{\rm esc}, E_0  \}\,.
    \end{split}
    \label{eq:astro_params}   
\end{equation}
Their fiducial values are taken from  the EOS2021 simulation~\cite{Munoz:2021psm}, and presented in table~\ref{tab:params}. The other parameters of the model which are not cited here are being fixed to their default value. We refer to refs.~\cite{Park:2018ljd, Qin:2020pdx,Qin:2020xyh,Munoz:2021psm} for a detailed description of the astrophysical model and parameters. 
AGC and MCG present different stellar-to-hallo mass relations (SHMR), parametrised by the fractions of galactic gas in stars $f_{*10,}^{II}$ and $f_{*,7}^{III}$ normalised to the value in halos of $10^{10} M_\odot$ and $10^7 M_\odot$ respectively. $\alpha_*^{II}$ and $\alpha_*^{III}$ are power law index controlling the mass dependence of the SHMR.
Similarly, the escape fractions (\textit{i.e} fraction of photons able to escape their host galaxy to ionize the IGM) are parameterised as a power law scaling with the mass of the halos, with  $f_{\rm esc,10}^{II}$ and $f_{\rm esc,7}^{III}$ the normalization of the ionizing UV escape fraction of high-z galaxies, evaluated for halos of $10^{10} M_\odot$ and $10^7 M_\odot$ respectively, and  with $\alpha_{\rm esc}$ the associated  mass scaling  power index.\footnote{We consider that ACG and MCG follow the same scaling with mass, so we only use one  common parameter~\cite{Munoz:2021psm}.}
Finally, $L_X^{II}$ and $L_X^{III}$ are the specific X-ray luminosity per unit star formation escaping the host galaxies, and determine the X-rays contribution to the heating and reionization of IGM. $E_0$ is the X-ray energy threshold below which photons can not escape the galaxy and do not contribute in the heating of the IGM.

\begin{table}[h]
    \centering
    \begin{tabular}{|c | c c c c c c c |c| }
        \hline
        Parameter & $h$                         & $\omega_b$    & $\omega_c$                  & $A_s \times 10^{-9}   $&  \multicolumn{2}{c}{$n_s$}         & \multicolumn{2}{c|}{$\tau_{\rm reio}$}\\
       \hline
        Fiducial  & 0.6736                    & 0.02237        &  0.1200                    &    2.1        &  \multicolumn{2}{c}{0.9649}        &   \multicolumn{2}{c|}{0.0544}\\
        Variation & 3\%                       &3\%             &3\%                         &3\%                     & \multicolumn{2}{c}{3\%}          & \multicolumn{2}{c|}{3\%}\\      
        \hline
        Parameter &$\log_{10}(f_{*,10}^{II})$ &$\alpha_*^{II}$ & $\log_{10}(f_{esc,10}^{II}) $ & $\log_{10}(L_{X}^{II})$ &  \multicolumn{2}{c}{$\alpha_{esc}$} & \multicolumn{2}{c|}{$E_0 $} \\
        \hline
        Fiducial  & -1.25                      & 0.5          &-1.35                         &40.5                   &      \multicolumn{2}{c}{ -0.3 }   &    \multicolumn{2}{c|}{ 500}     \\
        Variation & 3\%                        & 3\%          &3\%                          & 0.1\%                 &  \multicolumn{2}{c}{3\%}          &    \multicolumn{2}{c|}{3\%}\\
        \hline
        Parameter & $\log_{10}(f_{*,7}^{III}) $ &$\alpha_*^{III}$ &$\log_{10}(f_{esc,7}^{III})$  &  $\log_{10}(L_{X}^{III})$&  \multicolumn{2}{|c}{$ \Delta N_{\rm eff}$}&  \multicolumn{2}{|c|}{$N_{\rm idr}$}\\
        \hline
        Fiducial  & -2.5                      &  0.           &  -1.35                       & 40.5                          &  \multicolumn{2}{|c}{0} &   \multicolumn{2}{|c|}{0.} \\    
        Variation & 3\% &                  3\% $\alpha_*^{II}$               & 3\%                 & 0.1\%                  &  \multicolumn{2}{|c}{2.6\% $N_{\rm eff}$} & \multicolumn{2}{|c|}{ $\delta_{N_{\rm idr}}$}\\
        \hline
    \end{tabular}
    \caption{ Fiducial values of the cosmological and astrophysical parameters, as well as their variation for the computation of the numerical derivative in HERA Fisher matrix.
    Astrophysical parameters [see equation \eqref{eq:astro_params}] are only varied for the HERA Fisher matrix. Let us note that $\alpha_{*,\rm fid} ^{III} =0$, so we choose its variation according to $\alpha_*^{II}$, and that $\tau_{\rm reio}$ is varied only when we consider the CMB observations. The latter parameter is extracted from reionization history when combining 21cm and CMB information. Finally, we also vary $N_{\rm eff}$  (in section~\ref{sec:neff_section}) or $N_{\rm idr}$ (in section~\ref{sec:idmdr_section}) depending on the analyzed cosmological model. We refer to appendix~\ref{app:derivative} for a discussion about the value of $\delta_{N_{\rm idr}}$. The variation $\delta_{N_{\rm eff}} = 2.6 \% \cdot N_{\rm eff}$ has also been found following the method described in appendix~\ref{app:derivative}.}
    \label{tab:params}
\end{table}

\subsection{Fisher formalism}

To determine the sensitivity of HERA to astrophysical and cosmological parameters, we employ a \textit{Fisher matrix} analysis. For a list of parameters $\theta$, the components of the fisher matrix are written as 
\begin{equation}
    F_{ij} =  - \langle \frac{\partial^2 \ln \mathcal{L}}{\partial \theta_i \partial \theta_j} \rangle \,,
\end{equation}
where $\mathcal{L}$ is the likelihood distribution.
Assuming a Gaussian posterior distribution, the Fisher matrix corresponds to the inverse of the covariance matrix, while the error on a parameter $\theta_i$ follows $\sigma_i = \sqrt{F_{ii}^{-1}}$.  The Cramér-Rao theorem~\cite{frechet43,darmois45,aitken42} expresses that  the uncertainty computed on the parameter $\theta_i$ is the optimal result that can be obtained, given the configuration of the experiment for which we are making predictions.  
This approach is much faster than a  Monte-Carlo Markov Chain (MCMC) method. The downfall is that Fisher matrix analysis is less precise than MCMC and unreliable for non-Gaussian posteriors (see ref.~\cite{Mason:2022obt} for a comparison between the two methods on mock HERA 21cm-observations). As the bottleneck is the duration of a \texttt{21cmFAST} run,\footnote{A \texttt{21cmFAST} run takes around 1 hour on 4 cores for the configuration we have chosen.} we cannot rely on MCMC to perform our analysis. In this work, we perform ${\cal O}\sim 100$ of \texttt{21cmFAST} runs for the Fisher analysis,  two order of magnitude less than what would be needed for a MCMC analysis.

\subsubsection{21cm data}
We construct mock observations of the HERA telescope~\cite{DeBoer:2016tnn} which is designed to measure the 21cm power spectrum during Cosmic Dawn and the EoR.
The Fisher matrix  associated to the 21-cm power spectrum $\Delta^2_{21}(k,z)$ is given by: 
\begin{equation}
    F_{ij}^{21 \rm cm} = \sum_{k}\sum_{z}\frac{\partial \Delta_{21}^{2}(k,z)}{\partial \theta_i} \frac{ \partial \Delta_{21}^{2}(k,z)}{\partial \theta_j} \frac{1}{\sigma^{2}_{\Delta^{2}_{21} (k,z)}}\,,
\end{equation}
where $\sigma^2_{\Delta^2_{21}(k,z) }$is the measurement error in the 21cm power spectrum.
The $\Delta^2_{21}(k,z)$ power spectrum values are calculated from lightcones generated by \texttt{21cmFast} for a given set of parameters, and we refer to appendix~\ref{app:derivative} for a discussion on the numerical computation of the power spectrum derivatives $\partial \Delta_{21}^{2}(k,z) / \partial \theta_i$. We take boxes of size 256 Mpc, with a resolution of $128^3$ cells. In the computation of each matrix element, we sum over the scale bins $k$, and the redshift bins $z$. 
The measurement errors are evaluated with \texttt{21cmSense},\footnote{https://github.com/rasg-affiliates/21cmSense}~\cite{Pober:2012zz,Pober:2013jna}.

In this paper, we simulate a full HERA hexagonal array composed of 331 antennas of 14m diameter, located at a latitude of 30.8°. We assume an observation time of 6 hours per day for 540 days. The observations cover the frequency range 50 (z = 27.4) - 225 MHz (z= 5.3), with a bandwidth of 8 MHz corresponding to the redshift binning.  We only keep the bands corresponding to  redshifts $z>6$, because the signal is strongly suppressed below this limit due to reionisation leaving us with 19 frequency bands.  We adopt the \texttt{21cmSense} moderate foregrounds scenario,  assuming  that the foreground wedge extends up to  $a = 0.1 h Mpc^{-1}$ beyond the horizon wedge, and consider a HERA system temperature of $T_{\rm sys}(\nu) = T_{\rm receiver} + T_{\rm sky}(\nu) = 100K + 60 K (\nu/300 \rm MHz)^{-2.55}$. The associated thermal noise, scaling as $\propto T_{\rm sys}^2$, is linked to the beam, the defects, and the temperature of the antenna.
Finally, we limit the observable scales to  $ k =  0.1 - 1 \rm Mpc^{-1}$, as foreground increasingly contaminate large scale modes, and thermal noise drastically increase at large $k$.
The $\theta$ parameters varied in the Fisher matrix analysis are both of cosmological ($h$, $\Omega_b$, $\Omega_m$, $A_s$, $n_s$) and astrophysical origin (see eq.~\ref{eq:astro_params}). We will incorporate in the following the parameters of the extensions to $\Lambda$CDM considering in this paper, that we will detail in sections~\ref{sec:neff_section} and \ref{sec:idmdr_section}. The fiducial values of the parameters are listed in table~\ref{tab:params}.

\subsubsection{CMB data} 
In addition to characterising HERA constraining power on DM-DR interactions, our main goal is to establish the gain in sensitivity when combining 21cm observation with both present and future CMB information.

\paragraph{Planck} 
We first combined with Planck~\cite{Planck:2018vyg} which has already completed its observation program. We follow the procedure of~\cite{Lee:2022gz}, and take the inverse-covariance matrix \textbf{M} of the real data,\footnote{The data are publicly available at \url{https://pla.esac.esa.int/\#cosmology}} making the approximation that the data are Gaussian distributed. 

The fisher matrix is defined as: 
\begin{equation}
    F_{\theta_i, \theta_j}^{Planck} = \left(  \frac{\partial D_\ell}{\partial \theta_i} \cdot  \mathbf{M} \cdot \frac{\partial D_\ell}{\partial \theta_j} \right)\,,
\end{equation}
where $D_\ell = \ell(\ell +1) C_l /2\pi$ are the lensed temperature and polarization spectra ($D_\ell^{\rm TT}$, $D_\ell^{\rm TE}$,$D_\ell^{\rm EE}$).Let us note that we did not include lensing data since ref.~\cite{Archidiacono:2019wdp} has shown that the Planck lensing power spectrum has a very low impact on the constraints to the ETHOS model compared with the Planck primary power spectra.

\paragraph{Simons Observatory (SO)} 
We also perform forecast with  the next generation of ground-based observations from the CMB-SO telescope~\cite{SimonsObservatory:2018koc}. 
To calculate the fisher matrix associated with the measurements of this experiment, we use the same formulation as the one proposed for the CMB-S4  Fisher matrices~\cite{Flitter:2023mjj,Wu:2014hta,Munoz:2016owz,Adi:2020qqf,Shmueli:2023box}, adapted to the technical characteristics of CMB-SO, from ref.\cite{SimonsObservatory:2018koc}. 
The fisher matrix can be written as: 
\begin{equation}
    F_{\theta_i,\theta_j}^{SO} = \sum_{\nu} \sum_\ell \frac{2 \ell +1}{2} f_{\rm sky} \rm Tr \left [ C_\ell ^{-1} \frac{\partial C_\ell}{\partial \theta_i} C_\ell^{-1}\frac{\partial C_\ell}{\partial \theta_j}\right]\,,
\end{equation}
where $f_{sky} = 0.4$ is the widest possible sky-fraction coverage by CMB-SO, and where the $C_\ell$'s are the observables of interest, namely the CMB anisotropies temperature (T) and polarization (E) power spectra:\footnote{Carrying out a SO forecast including the lensing reconstruction implies extending the data to very small scales, which would require an improvement of the non-linear modeling, as non-linear corrections become important for those surveys (see \textit{e.g.} the discussion in the context of recent ACT data \cite{Madhavacheril_2024}). We thus decided to leave this for future work. \\}

\begin{equation}
   C_\ell(\nu)=
    \begin{bmatrix}
    \Tilde{C}^{TT}_\ell(\nu) & C_\ell^{TE}(\nu)\\
    C_\ell^{TE}(\nu) & \Tilde{C}^{EE}_\ell(\nu) 
\end{bmatrix} \,.
\end{equation}
We compute these 3 observables with \texttt{CLASS}~\cite{Lesgourgues:2011re}, and add a noise contribution from CMB-SO to the tilde quantities $\Tilde{C}_l^{XX} = C_l^{XX} + N_l^{XX}$, where $N^{XX}_l$ are the noise power spectra given by: 
\begin{equation}
\begin{split}
    N_\ell^{TT}=& \Delta_T^2(\nu) \rm e^{\ell(\ell+1)\sigma_b^2}\,, \\
    N_\ell^{EE} = & 2 \times N_\ell^{TT}\,.
\end{split}
\end{equation}
The noise contribution is driven by the  temperature sensitivity $\Delta_T$ modulated by $\sigma_b = \theta_{FWHM}/\sqrt{8 \log 2}$, where $\theta^2_{FWHM}$ is the  the full-width-half-maximum given in radians, which depends of the frequency channel. We consider three bands observed with the Large Aperture Telescope (LAT),  corresponding to the 93, 145 and 225 GHz frequencies, with respective noises of $\Delta_T = 8.0$, 10.0  and  22.0  $\rm \mu K$-arcmin, and with respective resolutions of $\theta_{\rm FWHM} = 5.1, 2.2$, and $ 1.4$ arcmin.

\paragraph{Planck + Simons Observatory}

CMB-SO will not be able to probe the largest angular bands of the CMB angular power spectrum (since it is a ground-based experiment), implying that we combine, in the following, the CMB-SO information with that of the Planck satellite~\cite{Planck:2018vyg}.
The SO and Planck data sets are independent, so we can sum the Fisher matrices to combine their information. 
We consider the Planck data in the range $\ell\in [2;2508]$ for the temperature power spectrum, and $\ell \in [2;1996]$ for the polarisation spectrum. For $\ell \ge 30$ we use the Planck-lite foreground-marginalized binned spectra and covariance matrix, and for $\ell<30$ we adopt the compressed log-normal likelihood of Prince and Dunkley~\cite{Prince:2021fdv}. 
We truncate the SO  mock data, keeping only the interval $\ell \in [2000;3000]$, in order to avoid double counting the information.\footnote{Although the $2000<\ell<2500$ are also present in the Planck data, the error bars at these multipoles are sufficiently large that this does not impact the results} In the following, we will refer to this dataset as \textit{CMB Mix}, which corresponds to:
\begin{equation}
    F^{\rm CMB Mix}_{\theta_i \theta_j} = F^{\rm Planck}_{\theta_i \theta_j} + F^{\rm SO_{2000 <\ell<3000}}_{\theta_i, \theta_j} \,,
\end{equation}
where $\theta$ are the $\Lambda$CDM parameters and the additional cosmological parameters inherent to the model we consider here. \\

When we analyze CMB data alone, we let the 6 $\Lambda$CDM (+ $\Delta N_{\rm eff}$) vary around their fiducial values taken from Planck TT,TE,EE+lowE+lensing  (see table~\ref{tab:params}). On the other hand, \texttt{21cmFast} provides us with a reionization history (encapsulated in the density-weighted ionization fraction $\overline{x_{HII} \cdot (1+ \delta_b)}$), that we can integrate in order to obtain the associated $\tau_{\rm reio}$, based on ref.\cite{Shmueli:2023box} computation. Therefore, when combining the CMB observations with the 21cm observables, we inject $\tau_{\rm reio}$ from the 21cm observations to the CMB analysis, making it directly dependent to the astrophysics of EoR. \\

We emphasize that forecasts are often optimistic with regard to the analysis that will be achievable with real data. In particular, foreground removal and instrumental systematics in 21cm data will certainly be more complex than predicted in our mock datasets. Another assumption that is made in Fisher forecasts is having fiducial values for the varied parameters. Although we vary the parameters around their fiducials, it is not guaranteed that the covariance matrix (and the derived uncertainties) would be the same if the fiducial values were different. We leave this complementary analysis for future investigation.

\section{First study: effective number of relativistic species}
\label{sec:neff_section}
\subsection{A brief review of the $N_{\rm eff}\Lambda$CDM model}

Before considering the effect of an interacting dark radiation component, it is instructive to start our analysis with a non-interacting one, as commonly parametrized by the {\it effective number of relativistic species} $N_{\rm eff}$. This parameter is defined through the energy density of relativistic species $\rho_{\rm R}$ as
\begin{equation}
     \rho_{\rm R} = \frac{7}{8} N_{\rm eff} \left(\frac{T_{\rm R}}{T_\gamma}\right)^{4} {\rho_\gamma} \, , 
     \label{eq:Neff}
 \end{equation}
where $\rho_\gamma$ is the photon energy density, and where $T_{\rm R}$ is the temperature of the relativistic species, which is different from the photon temperature $T_{\gamma}$ since these two components decoupled in the very early universe. We recall that for standard model neutrinos the temperature ratio is well determined by $T_{\rm R} /T_\gamma = (4/11)^{1/3} $.
Equation~\eqref{eq:Neff} allows to quantify any deviation from the standard assumptions, in particular that there are three generations of neutrinos (and no other relativistic particles) that follow the Fermi–Dirac distribution function (in equilibrium). In the standard model, $N_{\rm eff} = 3.044$~\cite{Akita:2020szl, Froustey:2020mcq, Bennett:2020zkv}, and any deviation from this number would imply the existence of a new relativistic particle, such as a light sterile neutrino species or an axion-like particle.
Therefore, in the following, we define $\Delta N_{\rm eff}$ as $\Delta N_{\rm eff} = N_{\rm eff} - 3.044$.
Let us note that this first study does not involve any modification beyond the background cosmology compared to the standard $\Lambda$CDM model, insofar as we have not (yet) introduced any interaction between the relativistic species and the dark matter component.

\subsection{21 cm meets the effective number of relativistic species}

\begin{figure}[ht]
    \centering
    \includegraphics[width=0.8\linewidth]{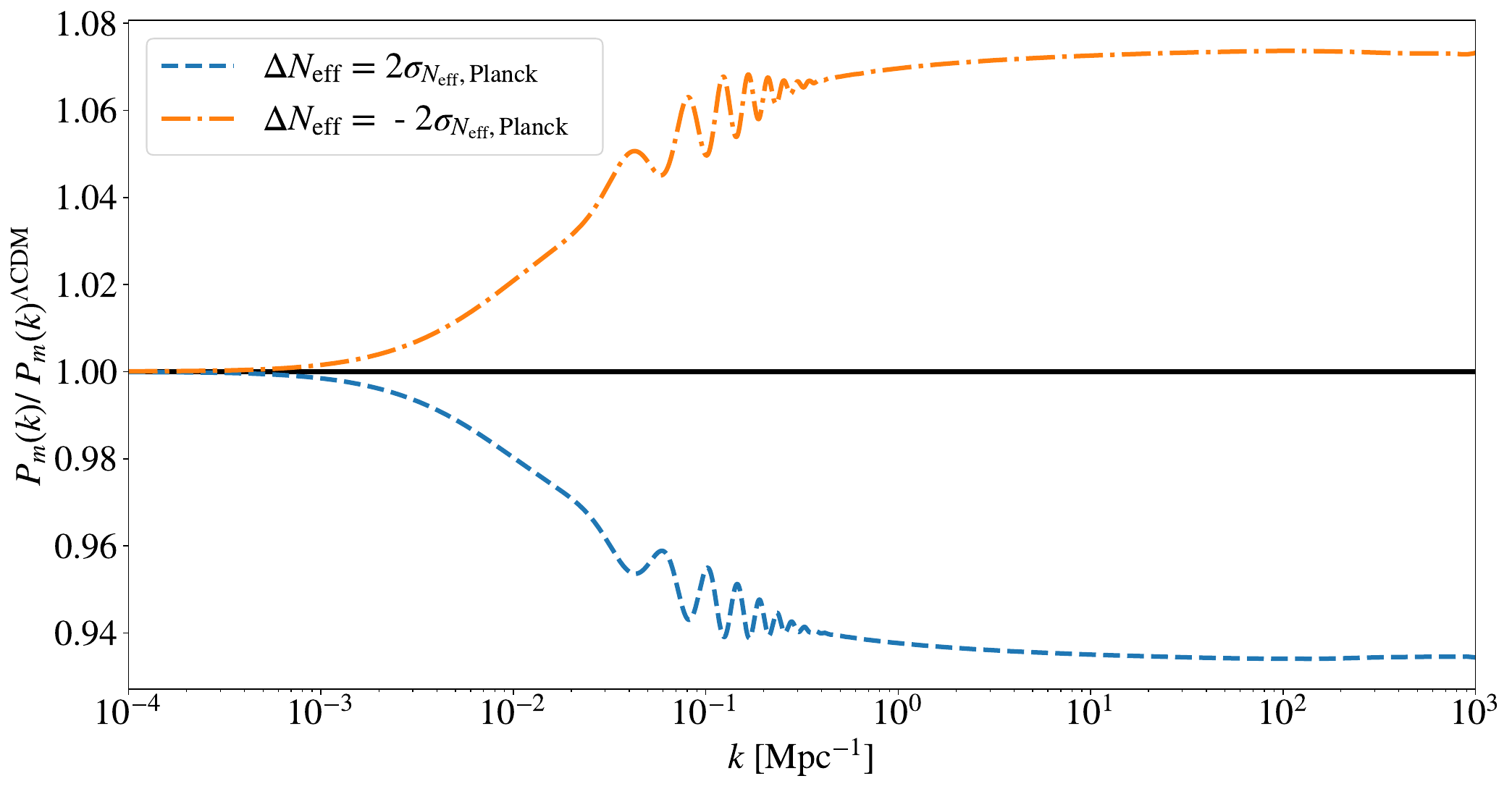}
    \caption{Residuals of the linear matter power spectrum $P_m(k)$ with respect to $\Lambda$CDM for the $ \pm 2 \sigma$ contour limit $\Delta N_{\rm eff} = 0.3536$  derived from our Planck fisher analysis (see table~\ref{tab:neff_cosmo_errors}). In this figure, we fixed the 6 $\Lambda$CDM parameters $\{ h, \Omega_m, \Omega_b, A_s, n_s, \tau_{\rm reio}\}$ to the Planck values~\cite{Planck:2018vyg}.}
    \label{fig:Pm_Neff}
\end{figure}

The effect of $N_{\rm eff}$ on the matter power spectrum is well known and is summarized in figure~\ref{fig:Pm_Neff}. First, an increase in this parameter leads to a decrease in the sound horizon at baryon drag  $r_s(z_{\rm drag})$, which induces a shift of the BAO towards small scales, explaining the wiggles in figure~\ref{fig:Pm_Neff}.
Second, the redshift of the matter-radiation equality is related to $N_{\rm eff}$ according to
\begin{equation}
    z_{\rm eq} = \frac{\omega_m}{\omega_\gamma \left[ 1 + \frac{7}{8} \left( \frac{4}{11} \right)^{4/3} N_{\rm eff} \right]} \, ,
\end{equation}
which implies that increasing $N_{\rm eff}$ decreases $z_{\rm eq}$.
Consequently, when all the $\Lambda$CDM parameters are fixed, increasing $N_{\rm eff}$ delays the onset of the matter domination era compared with the $\Lambda$CDM model. As a result, the growing mode solution of a sub-horizon perturbation has less time to grow, thus suppressing the small scales of the power spectrum, as can be seen in figure~\ref{fig:Pm_Neff}.\\

\begin{figure}[ht]
    \centering
    \includegraphics[width=0.8\linewidth]{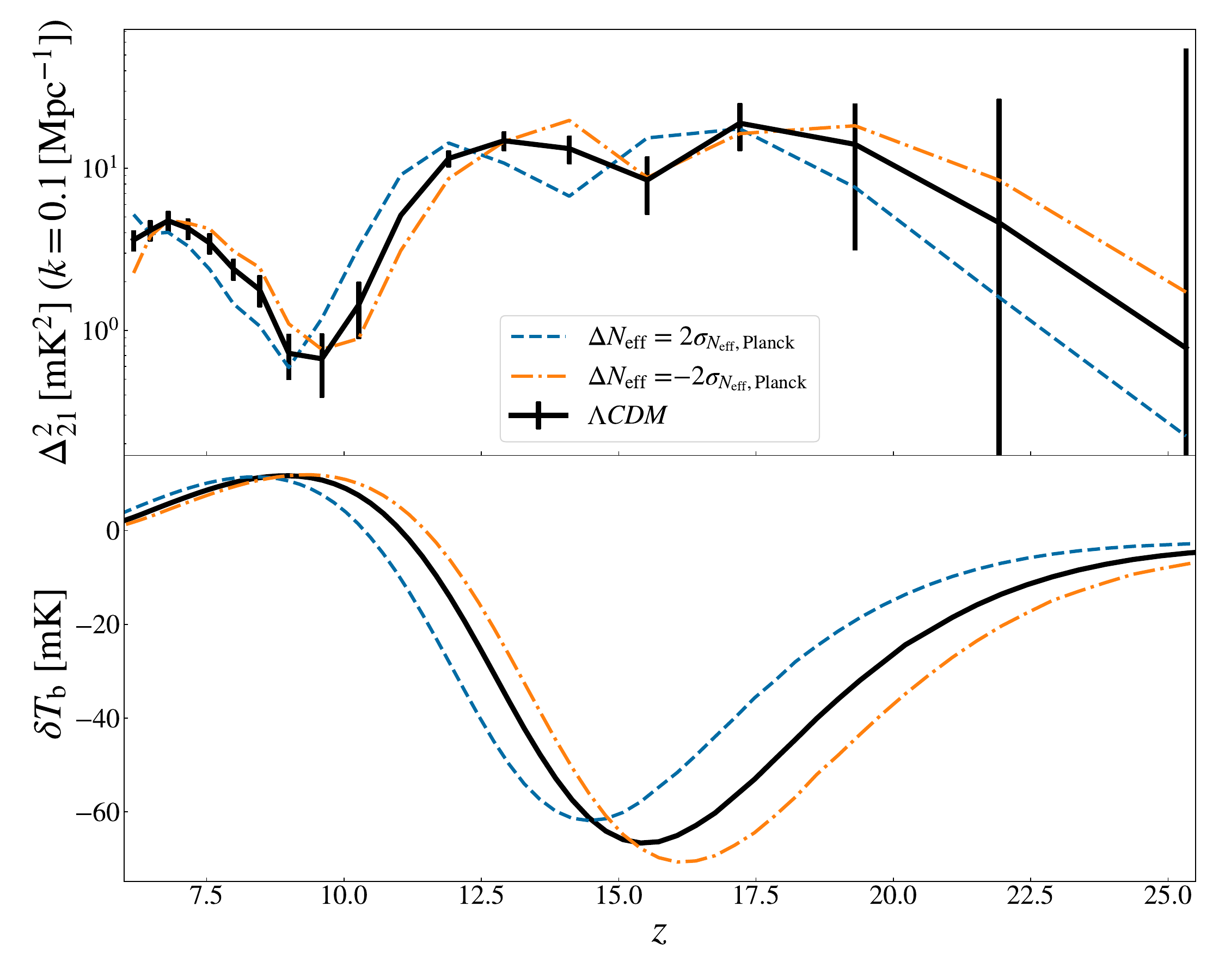}
    \caption{\textbf{Top:} 21cm power spectrum as a function of redshift at a fixed scale $k=0.1 \rm Mpc^{-1}$. We show it for the $ \pm 2 \sigma$ contour limits $\Delta N_{\rm eff} = 0.3536$ derived from our Planck fisher analysis (see table~\ref{tab:neff_cosmo_errors}). For comparison, we also displayed the fiducial $\Lambda$CDM cosmology from ref.~\cite{Planck:2018vyg}. Error bars are computed with \texttt{21cmSense}, assuming a moderate foreground scenario. \textbf{Bottom:} Same for the 21cm sky averaged differential brightness temperature. }
    \label{fig:neff_PS_GS}
\end{figure}

The effect of $N_{\rm eff}$ on the matter power spectrum will in turn have an effect on the 21cm signal, as shown in figure~\ref{fig:neff_PS_GS}. Our forecasts only account for the 21cm power spectrum, but we also present the global signal (in the bottom panel of figure~\ref{fig:neff_PS_GS}) which is a useful indicator of the effect of a model on  the 21cm signal. In general, the global signal presents a first feature in absorption after the formation of the first stars which produce UV photons interacting with neutral hydrogen gas through Wouthuysen-Field effect~\cite{wouthuysen52,field58}. In a second stage, X-ray photons heat the Intergalactic medium (IGM), driving the 21cm signal in emission. Finally, reionization progressively reduces the amount of neutral hydrogen, bringing the global signal back to zero. These three stages can also be observed in the large scales power spectrum (plotted in figure~\ref{fig:neff_PS_GS} as a function of the redshift at $k = 0.1 \rm  Mpc^{-1}$, the largest scale in our HERA mock observations expected to be relatively free from foreground), where three peaks are visible, reflecting respective domination of the three stages in the 21cm fluctuations. The troughs between them correspond to negative correlations of the different fluctuations, which cancel out for a time until one takes over the other in power. \\

If we consider the $N_{\rm eff} \Lambda$CDM model, with a positive $\Delta N_{\rm eff}$, the suppression of the small scales of the matter power spectrum are transferred in the halo mass function, inducing a reduction of the number of small mass halos with respect to $\Lambda$CDM. As a result, the formation of structures is delayed compared to the fiducial $\Lambda$CDM cosmology, deferring in turn the formation of the first stars, which produce the  Ly$\alpha$ flux leading to the absorption depth in the 21cm signal, as well as the $X$-ray heating of the gas. The 21cm features are therefore shifted to smaller redshifts, both in the global signal and 21cm power spectrum. Ly$\alpha$ pumping and X-ray heating epochs are more delayed than EoR, because those are the epochs at which MCG contribute the most. The latter, hosted in mini halos, are the most affected by $\Delta N_{\rm eff}$. As a result,  we also observe a reduction in amplitude  of the absorption feature in the  global signal, due to a faster evolution of the astrophysical epochs. For negative $\Delta N_{\rm eff}$, all the effects aforementioned  are reversed, and we observe a earlier onset of the Cosmic Dawn, associated with a general shift of the global signal and power spectrum to higher redshifts, as well as an increase of the absorption depth amplitude of the global signal.

\subsection{Results}

\begin{figure}[H]
     \centering
         \includegraphics[width=0.7\textwidth]{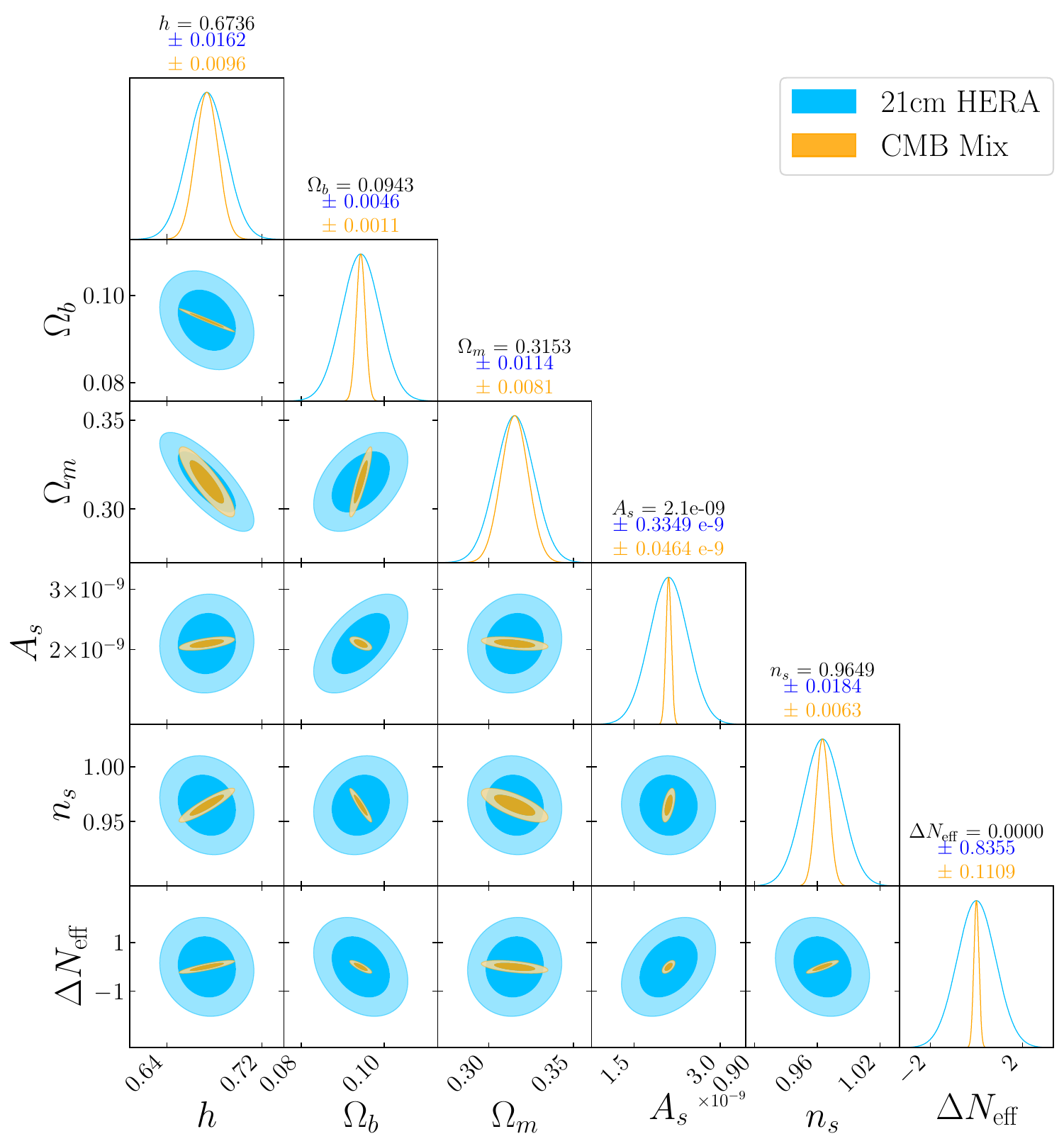}
         \caption{1-$\sigma$ and 2-$\sigma$ confidence level forecasts of the $ N_{ \rm eff}\Lambda$CDM model for the HERA 21cm and CMB Mix information. All the astrophysical parameters have been left free and marginalized in the calculation of the HERA fisher matrix, and results are shown for the moderate foreground scenario. Let us note that $\tau_{\rm reio}$ is varied in the CMB Mix fisher matrix and marginalized over.}
         \label{fig:ellipses_neff_21cm_cmbmix}
\end{figure}

\begin{figure}[ht]     
         \centering
         \includegraphics[width=0.7\textwidth]{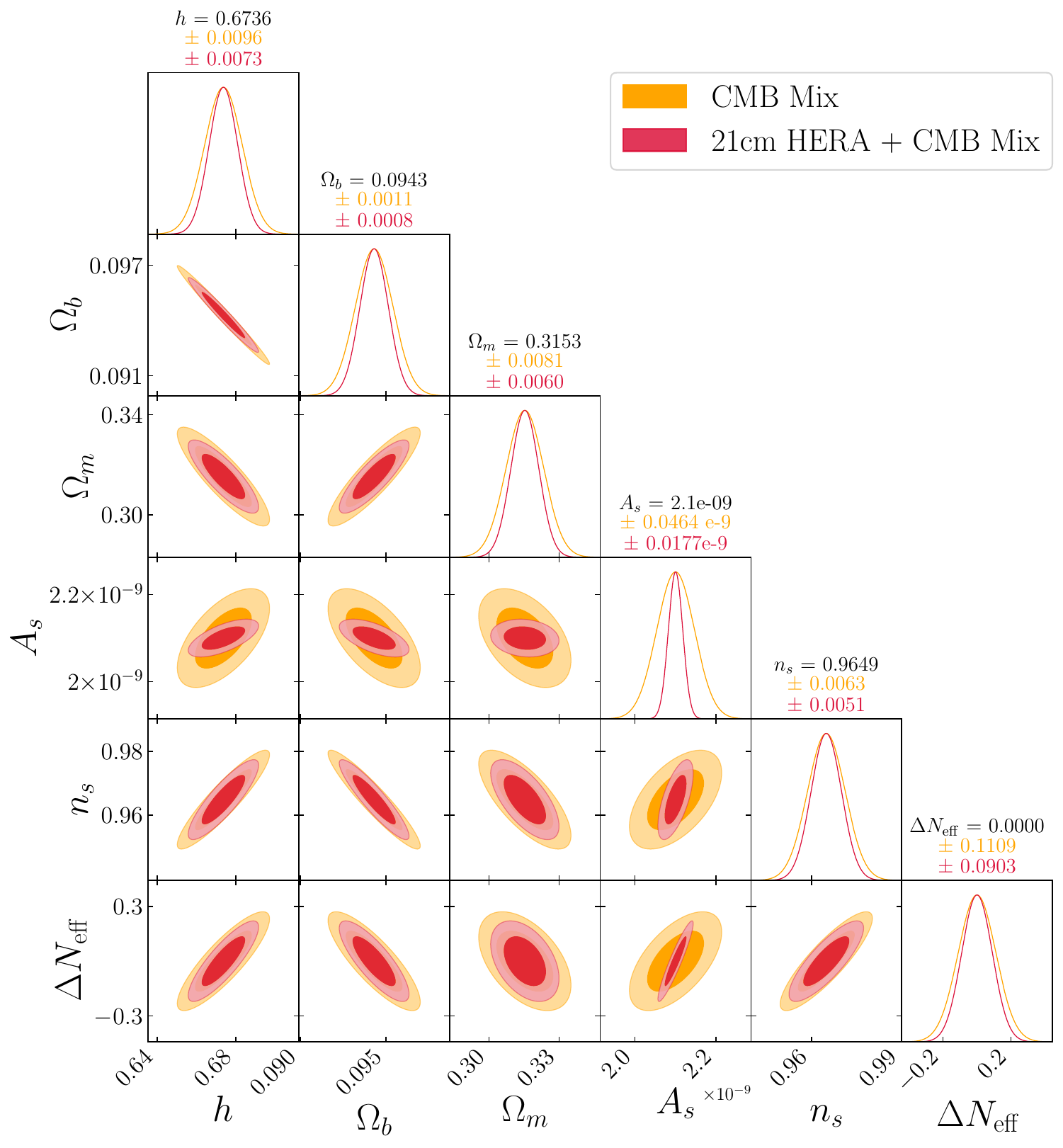}
     \caption{\textbf{Top:}  \textbf{Bottom:} Same as figure~\ref{fig:ellipses_neff_21cm_cmbmix} for the CMB Mix information and the combination between the HERA 21cm and CMB Mix information. When CMB is combined with HERA, the reionization history (and especially the $\tau_{\rm reio}$ parameter) is computed from \texttt{21cmFAST}.}
      \label{fig:ellipses_neff_cmbmix_21cmcmbmix}
\end{figure}

\begin{figure}[h]
    \centering
    \includegraphics[width=0.7\linewidth]{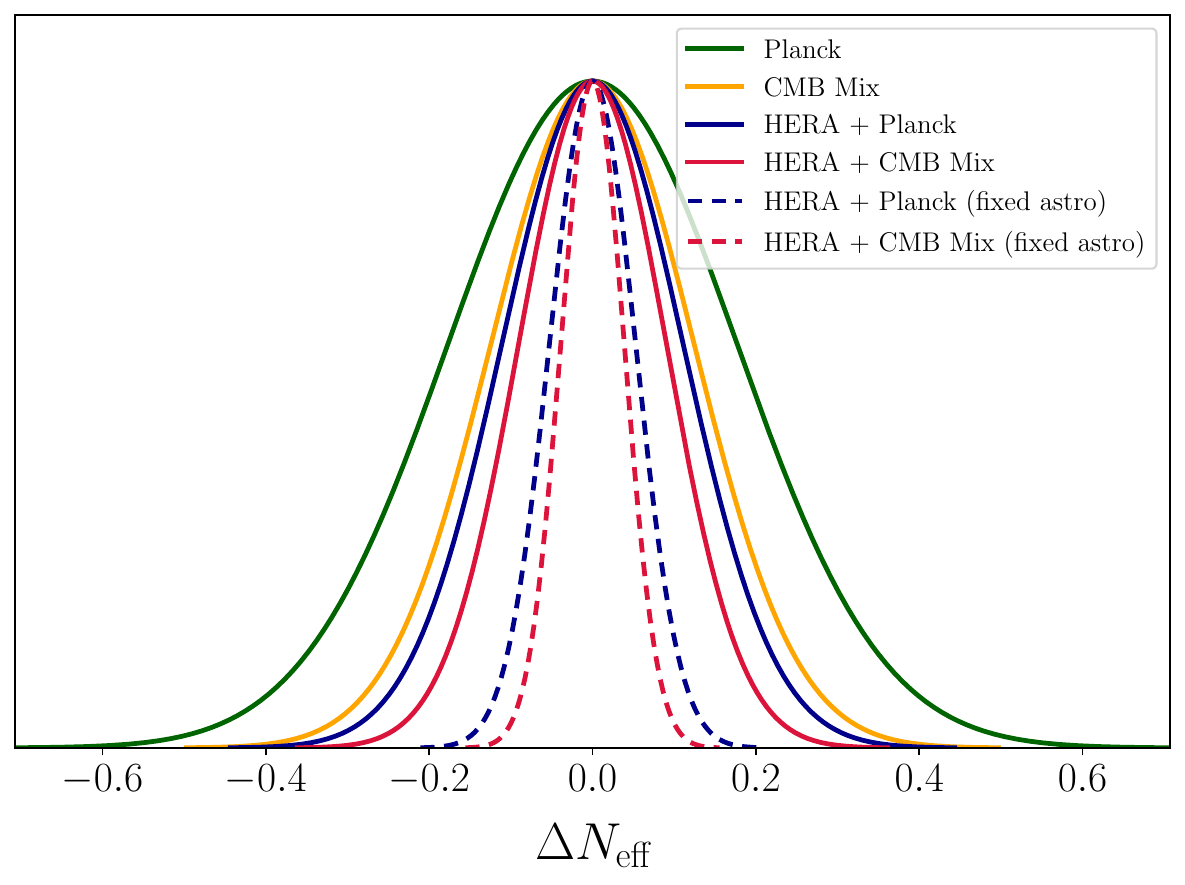}
    \caption{ 1-$\sigma$ confidence level forecasts of the $\Delta N_{\rm eff}$ parameter for Planck, CMB Mix, HERA + Planck and HERA + CMB Mix. For the last two analyses, we show the case where we vary both astrophysical and cosmological parameters, and the case where we fix the astrophysical parameters and where we vary only the cosmological parameters.}
    \label{fig:neff_1d_contour}
\end{figure}

We forecast the sensitivity of HERA to $\Delta N_{\rm eff}\neq 0$, and the additional information that can be provided by combining it with CMB data (Planck alone, and Planck + CMB SO).  Results from our analyses are shown in figures~\ref{fig:ellipses_neff_21cm_cmbmix}, \ref{fig:ellipses_neff_cmbmix_21cmcmbmix}, and~\ref{fig:neff_1d_contour}, while we provide in table~\ref{tab:params} the fiducial value of the parameters employed in our analysis, and the variation we apply to them to calculate the derivative of the 21cm power spectrum (see appendix~\ref{app:derivative} for more details). The estimated 1-$\sigma$ errors on the cosmological parameters can be found in table~\ref{tab:neff_cosmo_errors}. Finally, in figure~\ref{fig:ellipses_neff_fullastro} of appendix~\ref{app:neff_material}, we display the complete posteriors, including the astrophysical parameters. \\

In figure~\ref{fig:ellipses_neff_21cm_cmbmix}, we show the 1D and 2D marginalized posterior probability distributions from  the HERA and CMB Mix datasets. Let us note that for HERA the posteriors are marginalized over all the free astrophysical parameters of the model (see figure~\ref{fig:ellipses_neff_fullastro} of appendix~\ref{app:neff_material} for the posteriors of the astrophysical parameters). Our forecast suggest that HERA alone provides a poor constraining power on $\Delta N_{\rm eff}$ compared to CMB Mix, as the $1\sigma$ sensitivity of the former is weaker by factor of $\sim 7.5$  compared to the constraints of the latter. \\

In figure~\ref{fig:ellipses_neff_cmbmix_21cmcmbmix}, we compare the marginalized errors from CMB Mix alone with the combination of CMB Mix and  HERA.  
Interestingly, the combination of HERA and CMB data will improve the sensitivity to $N_{\rm eff}$, with estimated improvements of $42\%$ compared to Planck alone and $23\%$ compared to CMB Mix (see table~\ref{tab:neff_cosmo_errors}). 
In  addition, combining HERA and CMB information allows to significantly better constrain the parameter $A_s$, by a factor of $\sim 2.5$ comparing HERA + Planck with Planck alone, and by a factor $\sim 2.9$ comparing HERA + CMB Mix with CMB Mix alone.
This is due to our treatment of the optical depth to reionization, as described previously.
Given that $\tau_{\rm reio}$  and $A_s$ are strongly degenerate in the CMB (as they both scale the amplitude of the perturbations), by combining CMB information with HERA, we increase the information in the CMB fisher matrix, which explains the significant improvement in the constraints of $A_s$. 
This effect has been investigated by other papers, such as refs.~\cite{Liu:2015txa,Qin:2020xrg,Shmueli:2023box}. \\

Finally, in figure~\ref{fig:neff_1d_contour}, we compare the previous results with another scenario for 21cm observations, where we fix the astrophysical parameters to their fiducial value. As many efforts are pursued in order to improve our knowledge and constraints on the astrophysics during the CD and EoR, this represents the (very) ``optimistic'' scenario where the astrophysics is perfectly known. This is mostly intended for pedagogic purpose, to estimate what one could gain by improving our knowledge about astrophysical parameters. We refer for instance to ref.~\cite{Munoz:2024fas} for a discussion on the last constraints obtained from JWST on the ionizing photons, and to refs.~\cite{HERA:2021noe,HERA:2021bsv,HERA:2022wmy,Lazare:2023jkg} for the HERA extraction of an upper bound on the X-ray luminosity of the first stars. In this case, we expect to reach at best a $68\%$ limit of $\sigma_{\Delta N_{\rm eff}}$ = 0.0527 from HERA + Planck,  3.4 times better than the current constraints from Planck alone observations~\cite{Planck:2018vyg}  (see table~\ref{tab:neff_cosmo_errors}). We explore in more details the degeneracies between $\Delta N_{\rm eff}$ and the astrophysics parameters in appendix~\ref{app:neff_material}.\\

Summarizing, we find that HERA alone will not improve over the constraints from Planck, unless astrophysics is pinned-down from an independent experiment. Nevertheless, there is a strong complementarity between CMB and 21cm that will be interesting to exploit, in order to maximize the constraining power extracted from those datasets.\\

\begin{table}[!ht]
    \centering
    \begin{NiceTabular}{|c|c|c|c c|c c|c c|}
    \hline 
      &  Planck    &  CMB Mix  & \multicolumn{2}{|c|}{HERA} & \multicolumn{2}{|c|}{HERA + Planck}    &\multicolumn{2}{|c|}{HERA + CMB Mix}  \\
     \hline
      \diagbox{Param.}{Astro.}  & -   & -  & Fixed &  Free &  Fixed & Free & Fixed& Free  \\
        \hline
        $h$    &              0.0130 & 0.0096 &0.0133 & 0.0162  & 0.0066 & 0.0091 & 0.0057 & 0.0073\\
        $\Omega_b$ &          0.0015 & 0.0011 &0.0020 & 0.0046  & 0.0007 &0.0011  & 0.0006 & 0.0008\\
        $\Omega_m$&           0.0094 & 0.0081 &0.0105 & 0.0114  &0.0065  & 0.0069 & 0.0057 & 0.0060\\
        $A_s \times 10^{-9}$& 0.0510 & 0.0464 &0.1999 & 0.3349  & 0.0059 &0.0207  & 0.0056 & 0.0177 \\
        $n_s$              &  0.0077&  0.0063 &0.0106 & 0.0184  & 0.0037 & 0.0057 & 0.0032 & 0.0051 \\
        $\tau_{\rm reio}$   & 0.0112&  0.0104 & -     & -       & -      & -      &    -   &    -   \\
        $\Delta N_{\rm eff}$& 0.1768 & 0.1109& 0.5664& 0.8355  & 0.0527 & 0.1243 & 0.0389 & 0.0903 \\
        \hline
    \end{NiceTabular}
    \caption{Error bars at the $68\%$ confidence level for the $\Lambda$CDM cosmological parameters and $\Delta N_{\rm eff}$.  We show constraints from HERA in two situations, when we freely vary and when we fix the astrophysical parameters to their fiducial value in the HERA fisher matrix computation. When combined with HERA data, $\tau_{\rm reio}$ takes the values given by reionization history from  \texttt{21cmFAST} for the CMB fisher matrix computation.}
    \label{tab:neff_cosmo_errors}
\end{table}


\section{Second study: dark matter - dark radiation interaction}
\label{sec:idmdr_section}
\subsection{Review of the ETHOS model}

We now consider a model where the dark relativistic species interact with dark matter (iDMDR).
In this paper, we consider the ETHOS parametrisation, which has already been implemented in the \texttt{CLASS} code~\cite{Lesgourgues:2011re,Blas:2011rf} and allows to model a wide-variety of models of iDMDR. 
This kind of model requires the variation of (at least) two additional parameters compared to the $\Lambda$CDM model.
First of all, the impact of DR at the background level is quantified by the parameter $\xi = T_{DR}/T_{\gamma}$, which is the DR-to-photon temperature ratio, defined as
\begin{equation}
    \rho_{\rm DR} = \left(\frac{g_{\rm DR}}{2}\right) f_{\rm DR} \, \xi^4 \, \rho_\gamma \, ,
\end{equation}
where $f_{\rm DR}$ is a statistical factor equal to  $7/8$ for fermionic DR and to 1 for bosonic DR, and where $g_{\rm DR}$ is the DR number of internal degrees of freedom, assumed to be 2. Equivalently, we can use $N_{\rm idr}$, the effective number of DR species, defined as $\xi = (4/11)^{1/3} \times N_{\rm idr}^{1/4}$ in order to recover Eq.~\eqref{eq:Neff}.
Second, at the perturbation level, the interaction between DR and DM is quantified through the comoving scattering rate of DR off DM,
\begin{equation}
    \Gamma_{\rm DR-DM} = - \Omega_{\rm DM}h^2 a_{\rm dark}\left(\frac{1+z}{1+z_d}\right)^n \, ,
\end{equation}
where $z_d$ = $10^7$ is a normalisation factor, as well as the inverse scattering rate of DM off DR,
\begin{equation} \label{eq:Gamma_DMDR}
    \Gamma_{\rm DM-DR} = \left(\frac{4}{3}\frac{\rho_{\rm DR}}{\rho_{\rm DM}}\right) \Gamma_{\rm DR-DM} \, .
\end{equation}
The ``drag rate'' between DM and DR is thus controlled by two parameters: $n$ which is the temperature dependence of the interaction rate between dark matter and dark radiation, and $a_{\rm dark}$ which is the amplitude of the scattering rate.
The inclusion of a drag between DM and DR requires modifying the hierarchy of Boltzmann equations for the DR, written in Newtonian gauge as
\begin{align}
    \dot{\delta}_{DR} + \frac{4}{3}\theta_{DR} - 4 \dot{\phi} &= 0 \, , \label{eq:DR_perturb_1} \\
    \dot{\theta}_{DR} + k^2 \left(\sigma_{DR} - \frac{1}{4}\delta_{DR}\right) - k^2\Psi &= \Gamma_{DR-DM}\left(\theta_{DR} - \theta_{DM}\right) \, , \label{eq:DR_perturb_2} \\
    \dot{\pi}_{DR,\ell} + \frac{k}{2\ell+1}\left[(l+1)\pi_{DR,\ell+1} - \ell\pi_{DR,\ell-1}\right] &= \left(\alpha_\ell \Gamma_{DR-DM} + \beta_\ell \Gamma_{DR-DR}\right) \pi_{DR,\ell} \, , \label{eq:DR_perturb_3}
\end{align}
where $\delta$ is the density perturbation, $\theta$ the velocity dispersion perturbation, and  $\pi_{DR} = 2\sigma_{DR}$ is the shear perturbation. Finally, $\psi$ and $\phi$ are the gravitational potentials in Newtonian gauge.
In the last line, $\ell \in [2, \,  \ell_{\rm dark}]$, where $\ell_{dark}$ is the maximum multipole considered in the DR hierarchy, while  $\alpha_\ell$ and $\beta_\ell$ contain the interaction angular coefficients at order $\ell$ for DM-DR and DR-DR interactions respectively.
Let us note that in order to probe the very small scales, as required by the \texttt{21cmFAST} code (namely, up to $k = 1200 \,  {\rm Mpc}^{-1}$), we use the tight-coupling approximation of ref.~\cite{Archidiacono:2019wdp}.
Similarly, energy-momentum conservation requires to follow modified Euler perturbation equations for DM:
\begin{align}
                \dot{\theta}_{\rm DM} - k^2c^2_{\rm DM}\delta_{\rm DM} + \mathcal{H}\theta_{DM} - k^2 \Psi &= \Gamma_{\rm DM-DR} \left(\theta_{\rm DM} - \theta_{DR}\right) \, ,
    \label{eq:DM_perturb}
\end{align}
where $c^2_{\rm DM}$ is the dark matter sound speed. Let us stress that, when setting $a_{\rm dark} = 0 \, \rm Mpc^{-1}$, this model reduces to the $N_{\rm eff}$ model studied previously.\\

\subsection{Case study: 21 cm meets ETHOS $n=0$ model}

It is very common to fix $n$ in the statistical analyses as it is not a continuously varying variable, but rather quantifies the microscopic details of  the interaction (and in particular what kind of mediator is involved, see \textit{e.g}, ref.~\cite{Archidiacono:2019wdp}). Therefore, in our analyses, we vary only $a_{\rm dark}$ and $\xi$.
In this paper, we consider only the case $n=0$, which exhibits the phenomenology of various particle physics models, for which the DR-DM comoving scattering rate $\Gamma_{\rm DR-DM}$ is constant in time (and in temperature). This is for instance the case in non-Abelian DM models with strong self-interaction processes in the DR sector, which has been shown to be promising in resolving the $S_8$ tension~\cite{Buen-Abad:2015ova,Buen-Abad:2017gxg,Lesgourgues:2015wza,Krall:2017xcw,Pan:2018zha}.
For this model, it is necessary to include the effect of DR self-interactions, quantified by $\Gamma_{\rm DR-DR}$. 
In practice, we follow ref.~\cite{Archidiacono:2019wdp} and assume that DR is a relativistic perfect fluid (namely, $\Gamma_{DR-DR} \gg {\cal H}$), implying that we only need to consider the (perturbed) continuity and Euler equations, namely Eqs.~\eqref{eq:DR_perturb_1} and~\eqref{eq:DR_perturb_2}, and neglect Eq.~\eqref{eq:DR_perturb_3}.

\subsubsection{Effect on the Matter and CMB power spectra}

\begin{figure}[ht]
     \centering
     \begin{subfigure}[c]{0.48\textwidth}
         \centering
         \includegraphics[width=\textwidth]{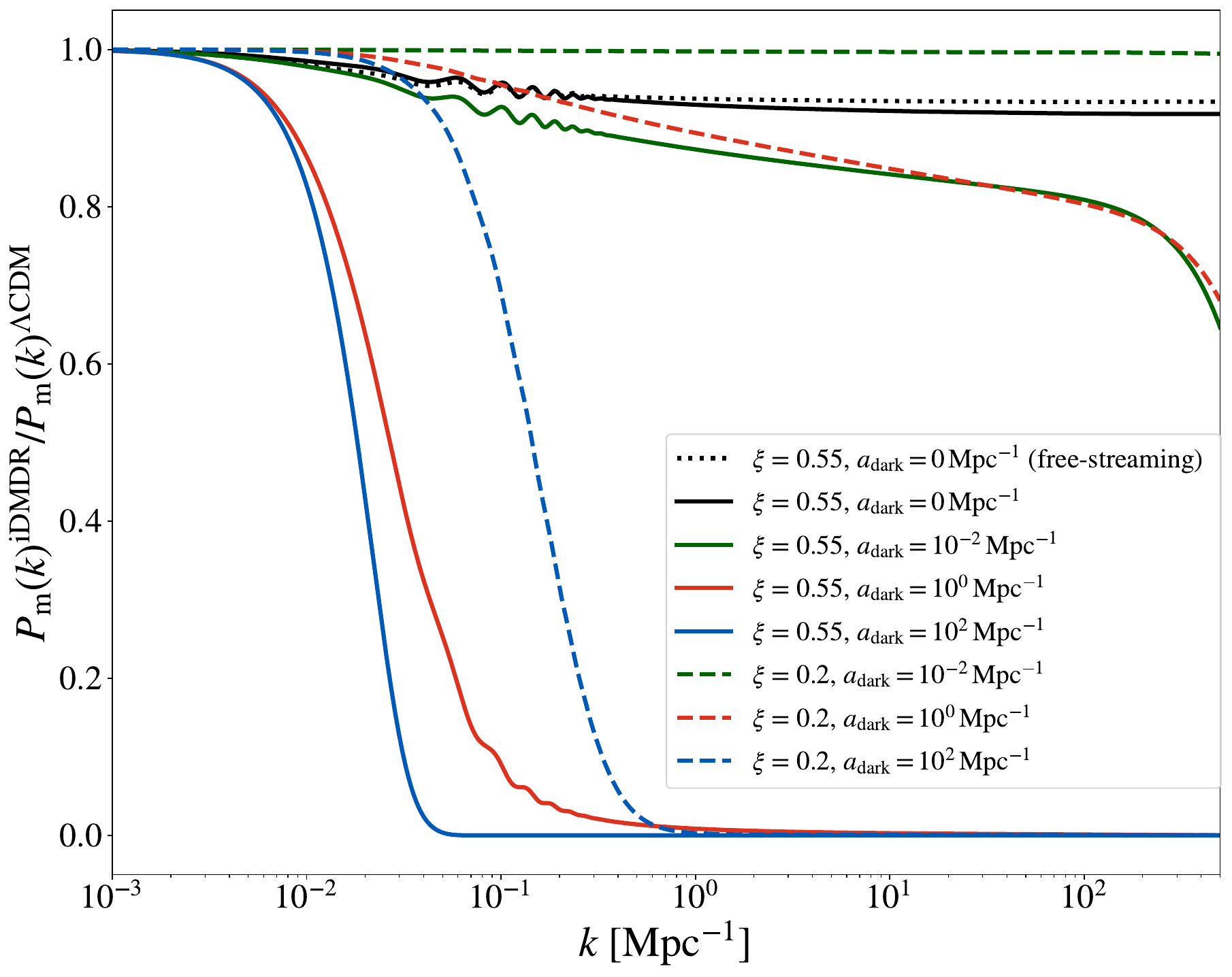}
     \end{subfigure}
    \hfill
     \begin{subfigure}[c]{0.50\textwidth}
         \centering
         \includegraphics[width=\textwidth]{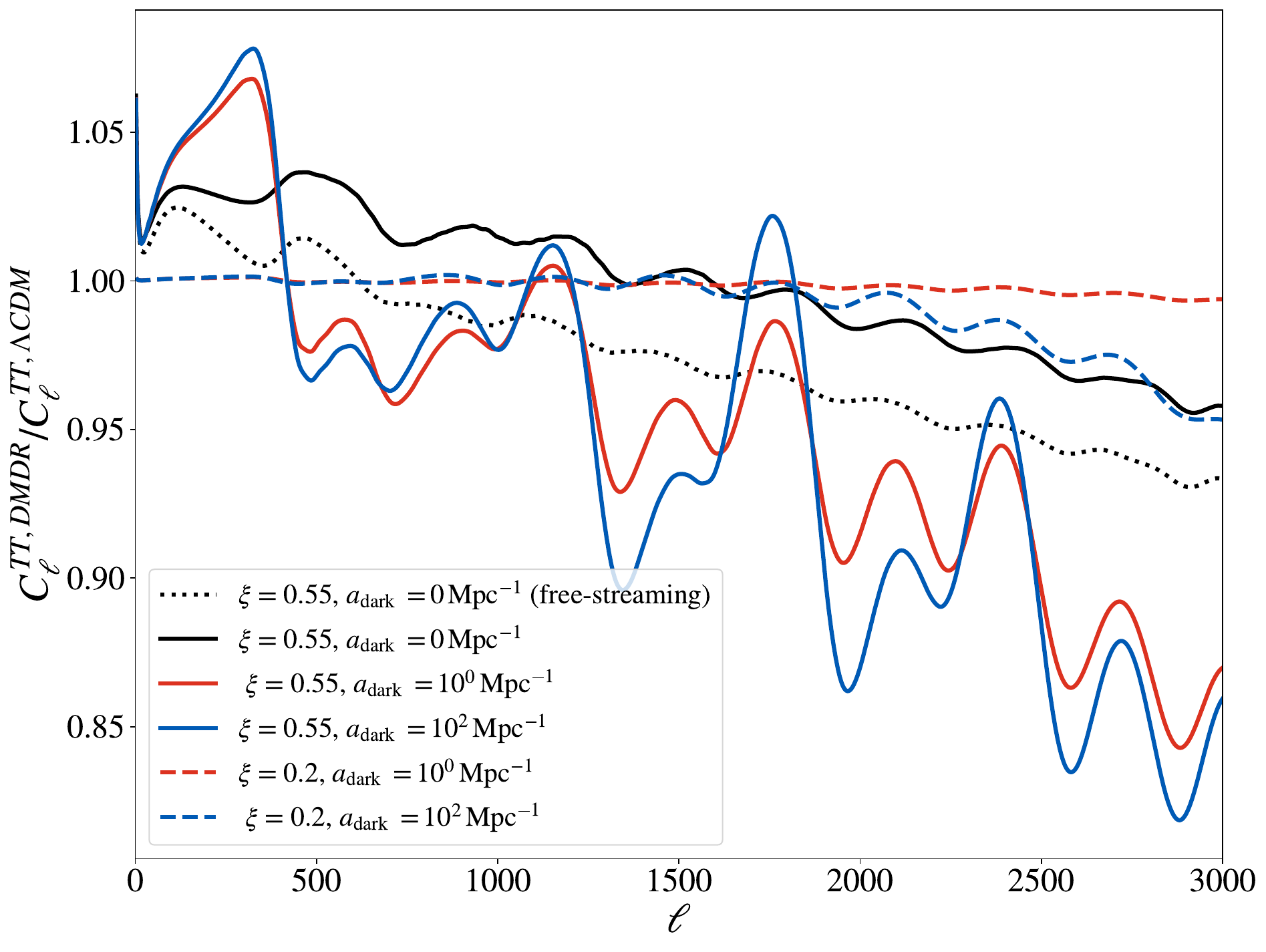}
     \end{subfigure}
        \caption{\textbf{Left:} Residuals of  the iDMDR matter power spectrum with respect to the $\Lambda$CDM fiducial cosmology (from ref.~\cite{Planck:2018vyg}), at $z=0$, for n=0, and for different values of $a_{\rm dark}$ and $\xi$.   The dotted black line is equivalent to the $\Delta N_{\rm eff} = 2\sigma_{N_{\rm eff},Planck}$ case presented in figure~\ref{fig:Pm_Neff}.  The red diagonally hatched area represents the scales accessible by  Planck CMB observations, while the blue horizontally hatched area illustrates HERA observable scales in our mock observations. Let us note that $\{ h, \omega_m, \omega_b, A_s, n_s, \tau_{\rm reio}\}$ are kept fixed.
        \textbf{Right:} Residuals of the temperature CMB power spectrum with respect to the $\Lambda$CDM fiducial cosmology for n=0. The angular scale of the sound horizon $\theta_s$ is kept fixed instead of $h$.}
        \label{fig:idmdr_PS_Cl}
\end{figure}

To illustrate the effect of the DM-DR interaction in $n=0$ ETHOS parametrisation on the matter power spectrum and the CMB temperature power spectrum, we plot in figure~\ref{fig:idmdr_PS_Cl} the residuals of these observables, relative to the $\Lambda$CDM cosmology.  We fix the cosmological parameters $\{ h, \omega_m, \omega_b, A_s, n_s, \tau_{\rm reio}\}$ in the matter power spectrum (left), while we fix $\{ \theta_s, \omega_m, \omega_b, A_s, n_s, \tau_{\rm reio}\}$ in the  CMB temperature power spectrum (right). Let us note that $\xi$ = 0.55 can be converted in $N_{idr} = 0.3536$, in order to be compared with the $\Delta N_{\rm eff}$ model. 
In the limit where there is no interaction with dark matter ($a_{\rm dark} = 0 \, \rm Mpc^{-1}$), the presence of dark radiation has the same background effect on the matter and CMB power spectra as the $N_{\rm eff}\Lambda$CDM model. However, in the specific case of $n=0$, DR is treated as a perfect fluid because of its self-interaction. As a result, the decrease in amplitude and shift of the acoustic peaks to smaller scales that are expected from the `neutrino drag' effect of a free-streaming radiation $\Delta N_{\rm eff}$~\cite{particledatagroup22} are alleviated in presence of non free-streaming DR, as can be seen in the right panel of figure~\ref{fig:idmdr_PS_Cl} (black solid vs dotted lines). 
Approximating DR as a fluid also induces a slight variation on the matter power spectrum resulting from the absence of DR anisotropic stress compared to the free-streaming case, visible in the left panel of figure~\ref{fig:idmdr_PS_Cl}  (black solid vs dotted lines). \\

At the level of the perturbations, the DM-DR interaction acts against the gravitational collapse of the DM, which results in a stiff suppression of the matter power spectrum for the modes entering the horizon before the kinetic decoupling between DM and DR. In the left panel of figure~\ref{fig:idmdr_PS_Cl}, we can indeed see that this suppression is enhanced as we increase the interaction strength $a_{\rm dark}$. Let us note that the comoving interaction rate acting in the DM Euler equation (see eq.~\ref{eq:DM_perturb}) shows a $\Gamma_{\rm DM-DR} \propto a_{\rm dark} \xi^4$ dependence, implying that $\xi$ and $a_{\rm dark}$ are degenerated at the level of the perturbations.
In addition, the reduction in the DM clustering driven by the interaction is reflected as well in the baryon-photon fluid. This induces a suppression of the temperature CMB power spectrum at small scales, as shown in the right panel of figure~\ref{fig:idmdr_PS_Cl}. As a result, both $\xi$ and $a_{\rm dark}$ suppress this observable and the effect of these parameters on the CMB power spectrum are strongly degenerated as well. 
We refer to refs.~\cite{Cyr-Racine:2013fsa,Archidiacono:2017slj} for a more detailed discussion on the effects of DM and DR interaction on the CMB power spectrum.\\ 

Let us note that, in cases where the comoving interaction rate has a higher temperature dependence (typically, $n=2$ or $n=4$), the DM-DR interaction also leads to the production of dark acoustic oscillations (DAO) appearing in the matter power spectrum~\cite{Cyr-Racine:2013fsa,Bohr:2020yoe,Munoz:2020mue,Verwohlt:2024efh}. They arise from the opposition between the gravitational force from DM clustering, and the pressure force from DR. However, DAOs do not occur in the $n=0$ case we are studying here, because the interaction rate $\Gamma_{\rm DM-DR}$ has the same temperature dependence as the expansion rate during the radiation dominated era.

\subsubsection{$n=0$: 21cm Global Signal \& Power spectrum}

The suppression of the matter power spectrum at small scales induces a cut-off in the halo mass function (HMF),  resulting in a reduction of the low-mass halo population.
In this work we make use of a top-hat filter function in the computation of the HMF, which is the common choice when studying CDM. 
The top-hat filter function tends to overestimate the low-mass halo populations when confronted to models inducing a suppression in the power spectrum at small scales, such as warm dark matter or DM-DR interaction~\cite{Schneider:2013ria,Schneider:2014rda,Schaeffer:2021qwm,Benson:2012su,Leo:2018odn}. Efforts have been made to propose a more accurate filter function. 
For instance, the `sharp-$k$' filter presents an abrupt cut-off, and provides better results for models inducing a smooth suppression of the matter power spectrum~\cite{Schneider:2013ria}. Lately, to account for DAO features such as in the $n=2$ and $n=4$ cases of the ETHOS parametrisation, a `smooth-$k$' filter has been proposed~\cite{Schewtschenko:2014fca,Leo:2018odn, Schaeffer:2021qwm,Bohr:2021bdm}, controlled by parameters which are fitted from the results of N-body simulations. We leave the analysis with different filter functions for a future work. It should be pointed out that the suppression in the HMF from DM-DR interaction is underestimated due to our choice of top-hat filter function. Hence, our forecasts regarding HERA's sensitivity are likely on the conservative side. For the computation of the HMF itself, we use the Sheth-Tormen fitting function~\cite{sheth01}, which was calibrated based on CDM N-body simulations.\\

\begin{figure}[ht]
     \centering
     \begin{subfigure}[b]{0.48\textwidth}
         \centering       
        \includegraphics[width=\textwidth]{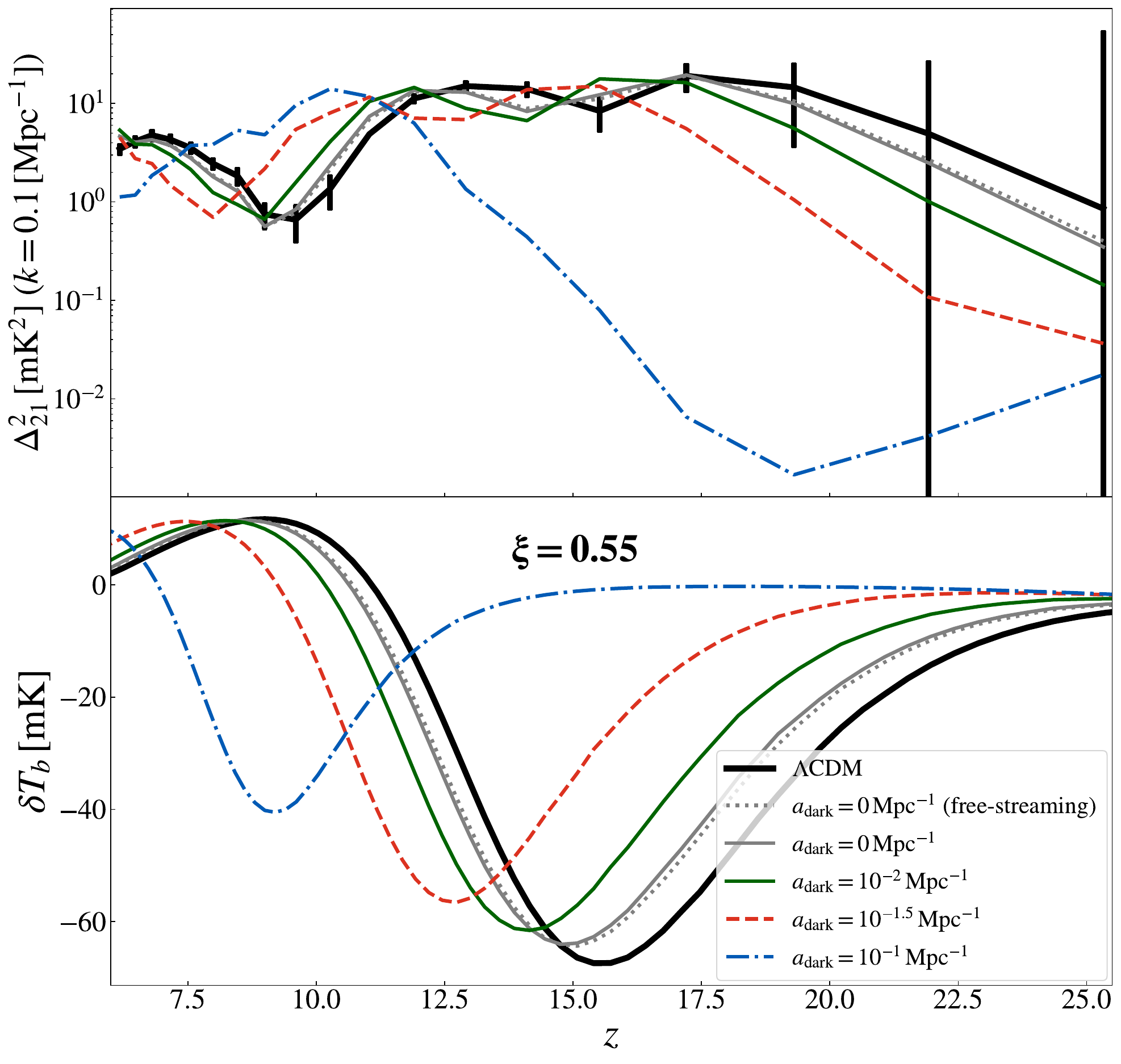}
     \end{subfigure}
     \begin{subfigure}[b]{0.49\textwidth}
         \centering
         \includegraphics[width=\textwidth]{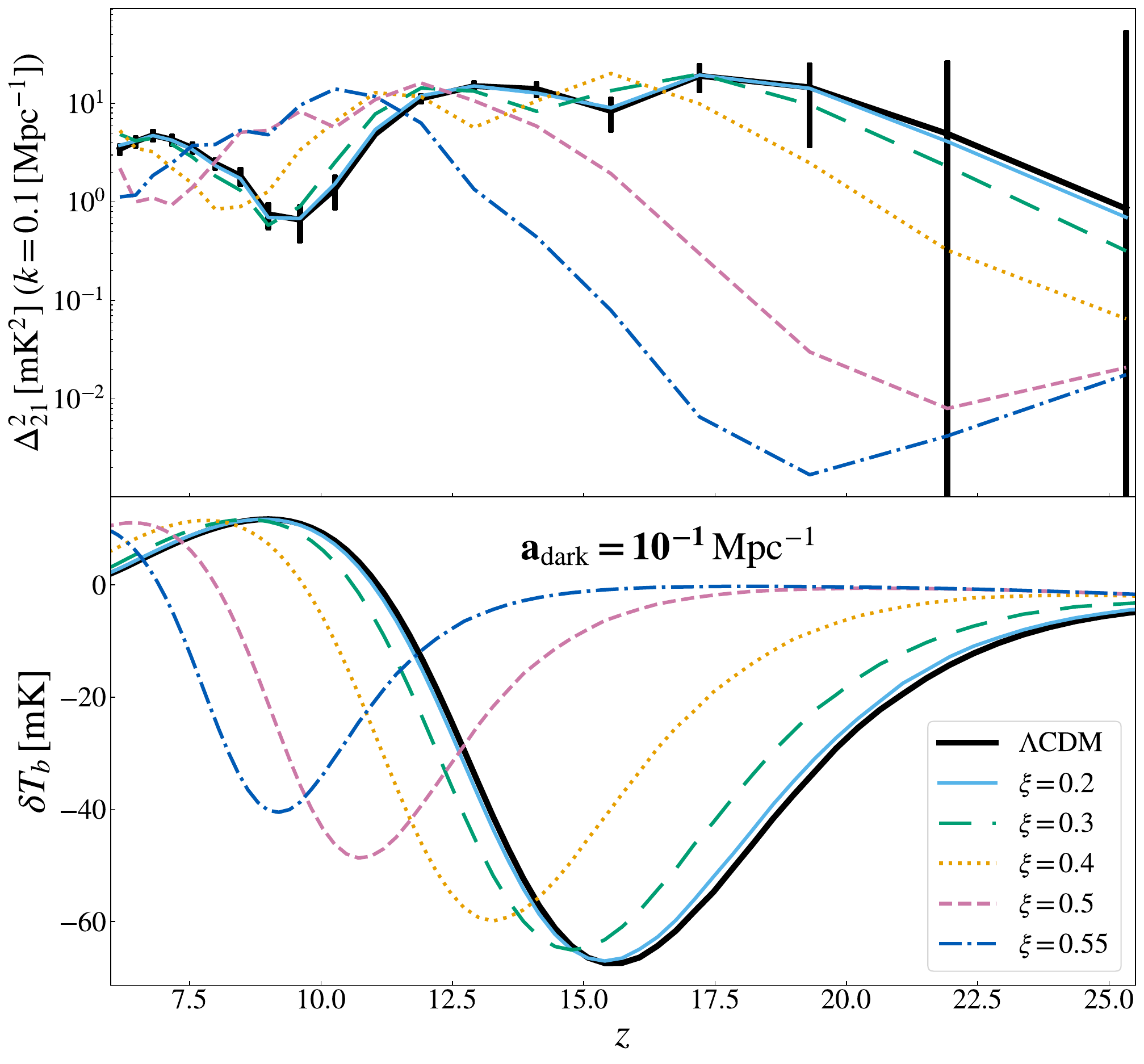} 
     \end{subfigure}
        \caption{\textbf{Top:} 21cm power spectrum as a function of redshift at a fixed scale $k=0.1 \rm Mpc^{-1}$. We show it for different values of $a_{\rm dark}$  fixing $\xi = 0.55$ (\textbf{Left}), and for different values of  $\xi$ fixing $a_{\rm dark} = 10^{-1} \, \rm Mpc^{-1}$ (\textbf{Right}). For comparison, we also display the fiducial $\Lambda$CDM cosmology from ref.~\cite{Planck:2018vyg}. Errorbars are computed with \texttt{21cmSense}, assuming a moderate foreground scenario. \textbf{Bottom:} Same for the 21cm sky averaged differential brightness temperature.}
        \label{fig:idmdr_GS_PS}
\end{figure}

We illustrate the impact of different DM-DR  scenarios on the 21cm signal in figure~\ref{fig:idmdr_GS_PS}.
On the left panel of this figure, we fix the amount of dark radiation to $\xi =0.55$ and vary the strength of the interaction $a_{\rm dark}$, while on the right panel, we do the opposite by fixing $a_{\rm dark}  = 10^{-1} \, \rm Mpc^{-1}$ and varying $\xi$. Since these two parameters induce a suppression on the matter power spectrum (as described previously), their impact on the 21cm signal is rather similar: when we increase the amount of dark radiation and/or the strength of the DM-DR interaction, we observe a shift of the 21cm signal toward the lower redshifts  due to the suppression of the low mass halo population in the HMF which delays the structure formation. \\

A similar behavior can be observed in the 21cm power spectra at $k = 0.1 \rm Mpc^{-1}$, as can be seen in the top panels of figure~\ref{fig:idmdr_GS_PS}: strong DM-DR interactions delay the three peaks (representing the domination of the fluctuations associated to Ly$\alpha$ flux, gas temperature, and ionization fraction) to lower redshifts.
Moreover, in the left panel of figure~\ref{fig:idmdr_GS_PS}, we first observe that, similarly to the effect on the matter power spectrum, the fluid approximation on DR induces a very weak variation of the 21cm signal compared to the free-streaming DR studied in section~\ref{sec:neff_section} (solid vs dotted grey lines).

\subsection{Results for the ETHOS $n=0$ model}

\begin{figure}[!h]
     \centering
     \begin{subfigure}[b]{0.49\textwidth}
         \centering
         \includegraphics[width=\textwidth]{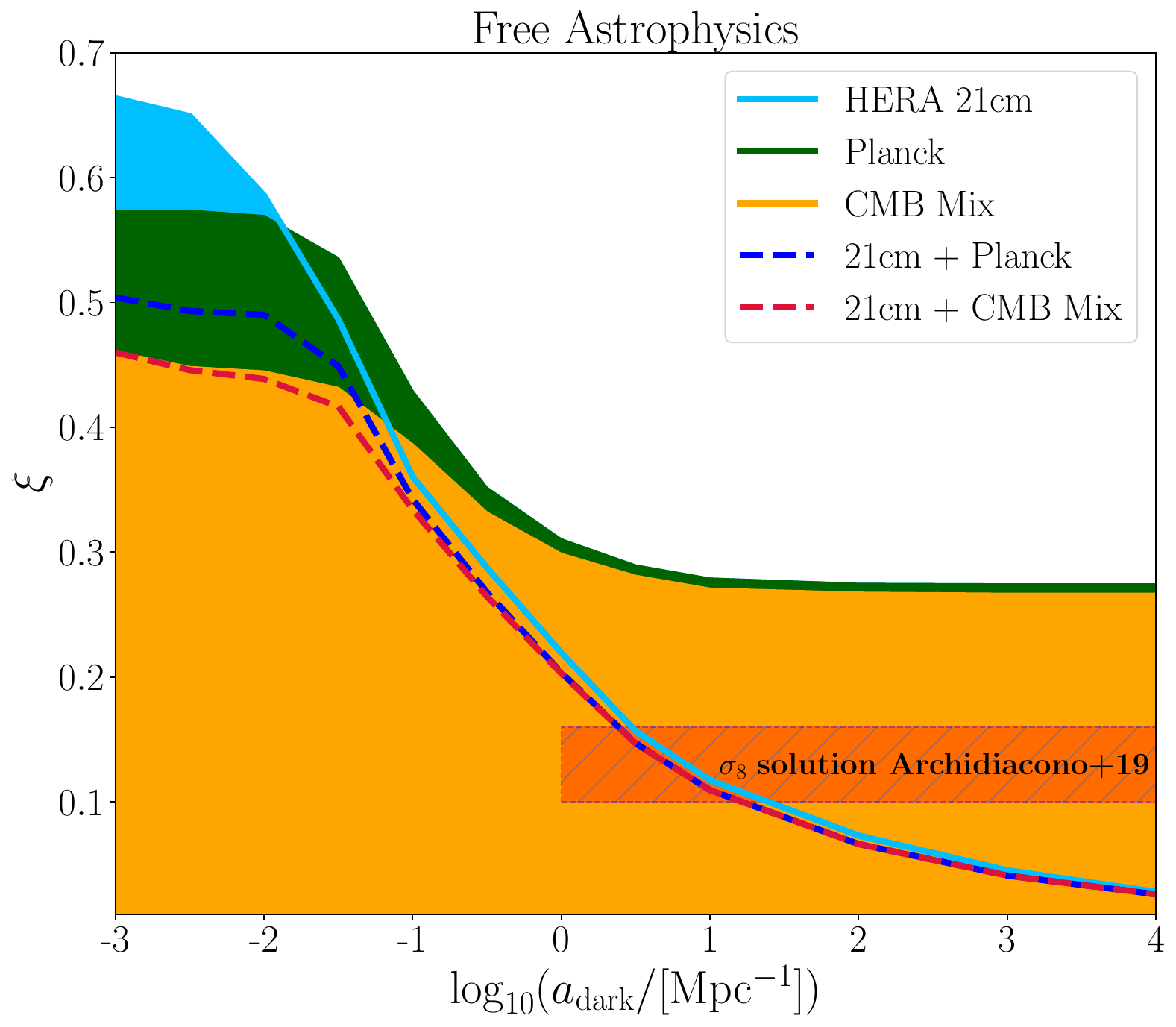}
     \end{subfigure}
     \begin{subfigure}[b]{0.49\textwidth}
         \centering
         \includegraphics[width=\textwidth]{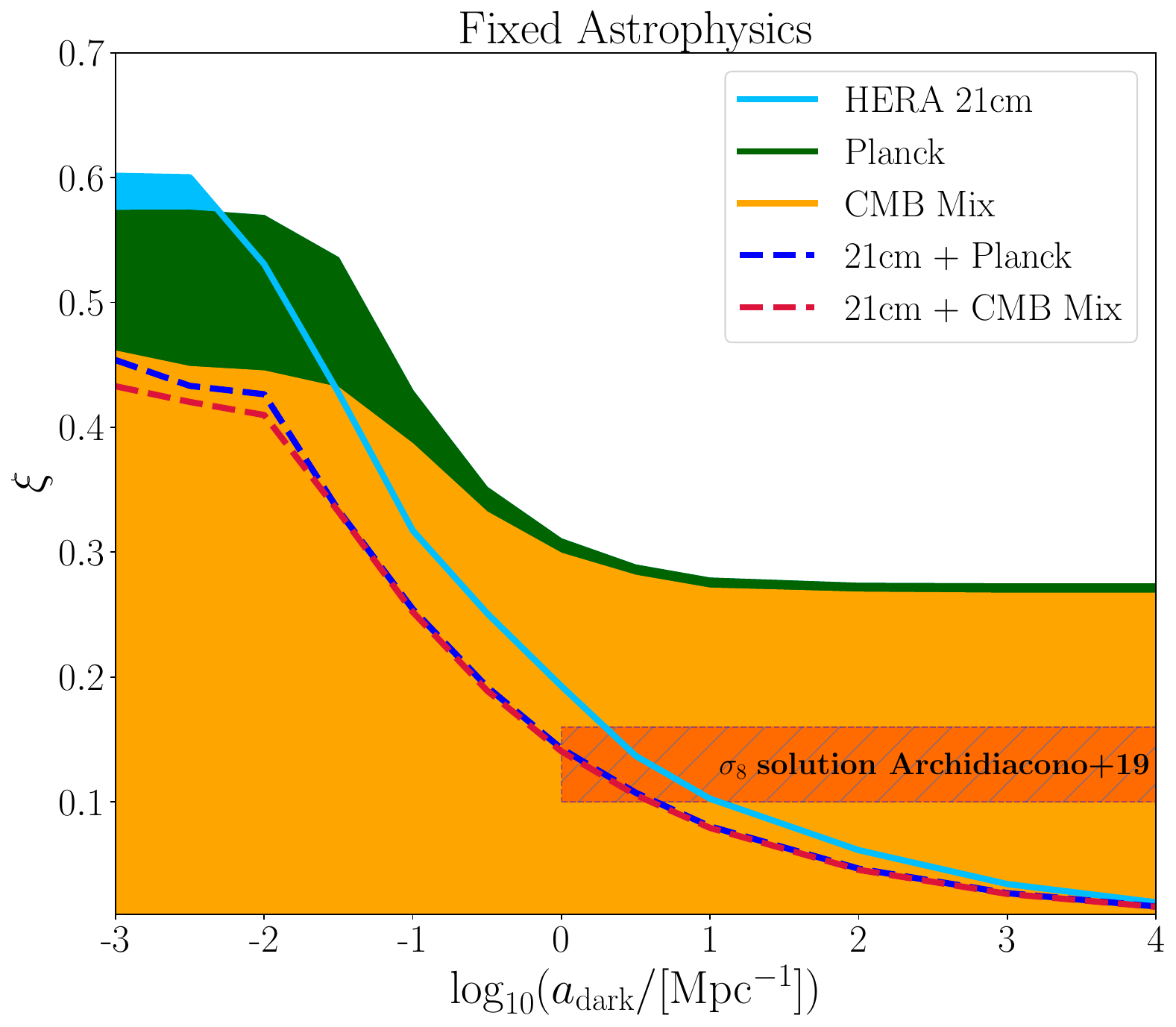}
     \end{subfigure}
     \begin{subfigure}[b]{0.49\textwidth}
         \centering
         \includegraphics[width=\textwidth]{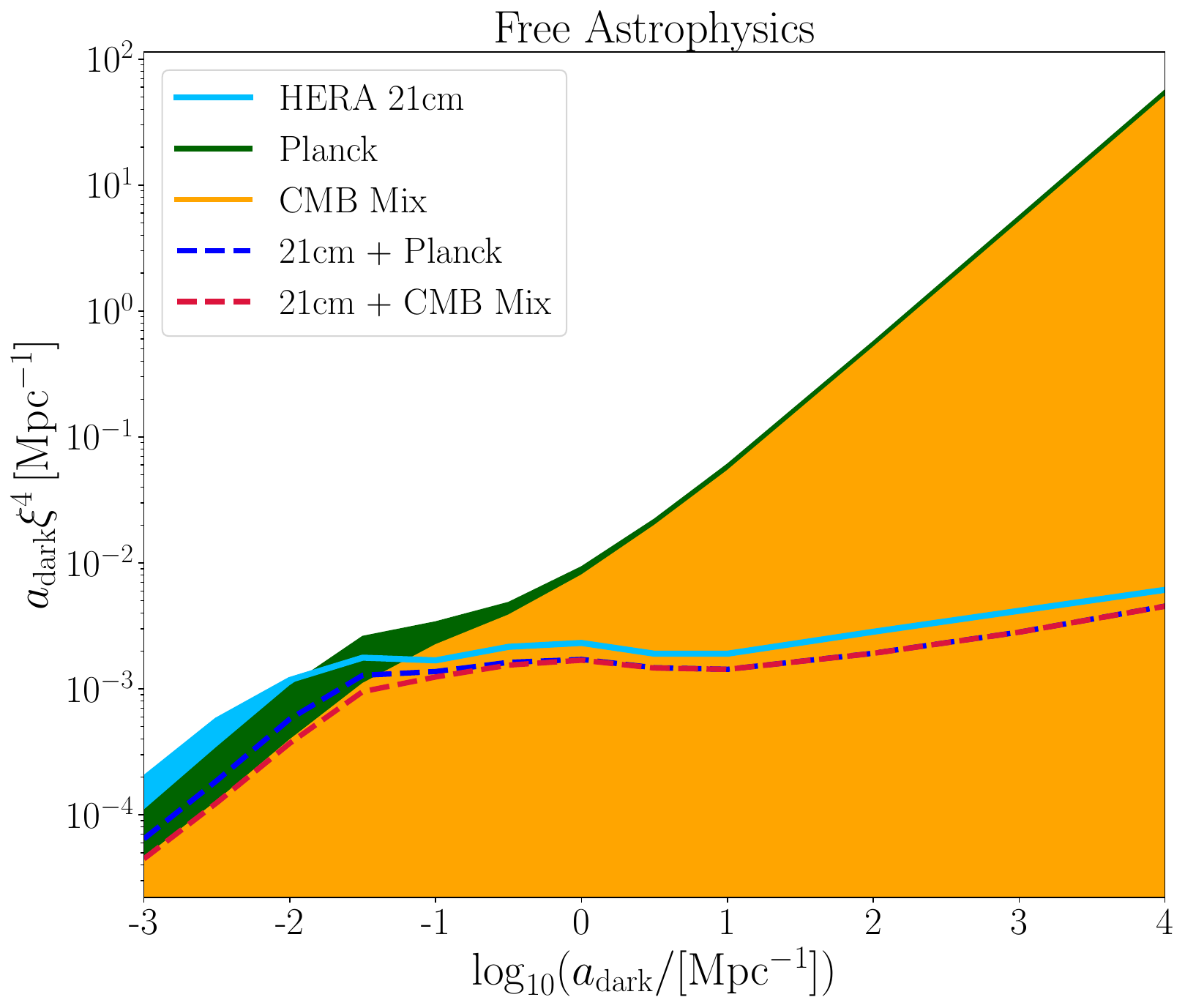}
     \end{subfigure}
     \begin{subfigure}[b]{0.49\textwidth}
         \centering
         \includegraphics[width=\textwidth]{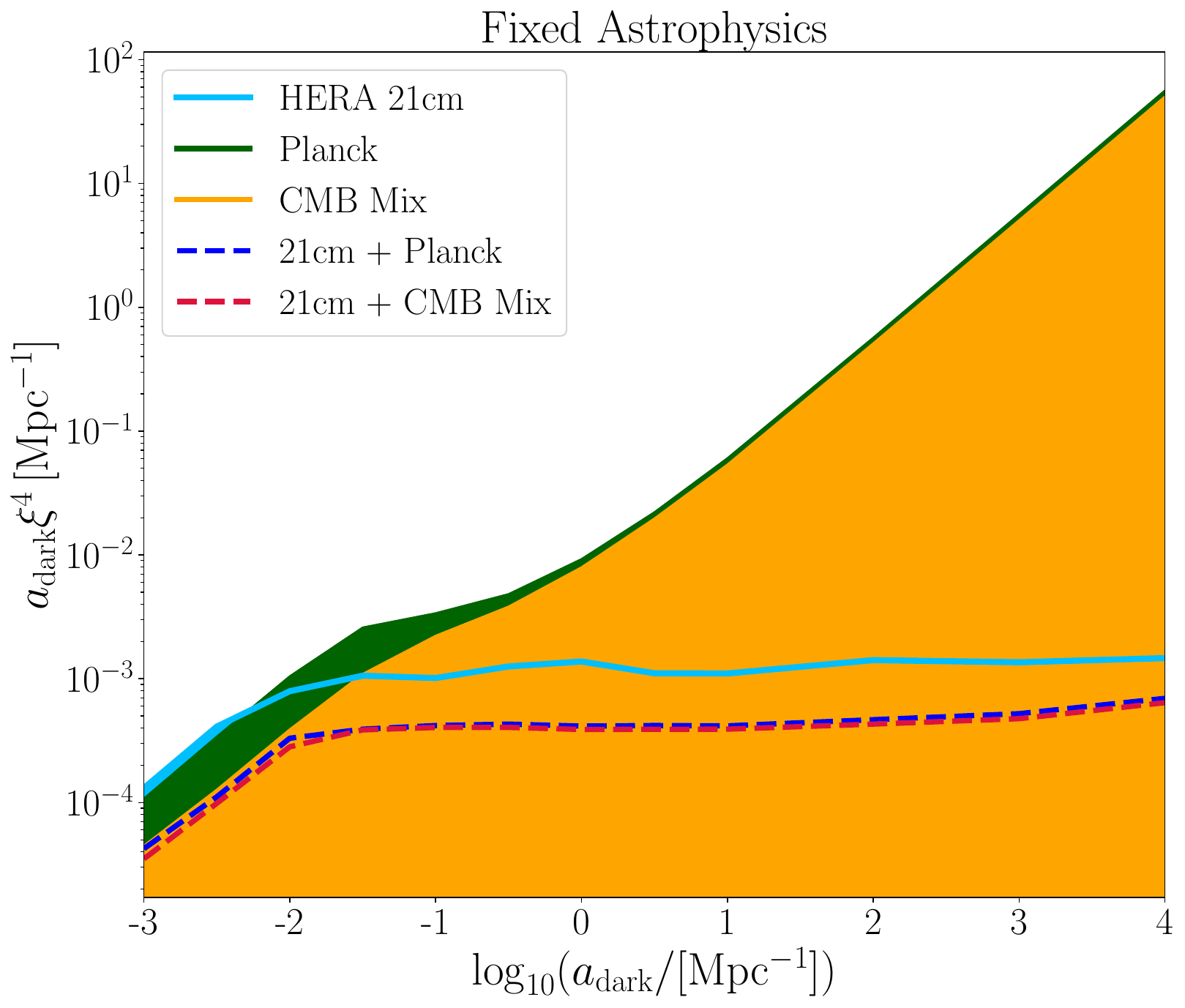}
     \end{subfigure}
     \caption{\textbf{Top:} Sensitivities (at 95\% C.L.) on the dark radiation energy density $\xi$, computed for different values of $a_{\rm dark}$. As for the $N_{\rm eff}\Lambda$CDM model, we show the HERA sensitivities for the free astrophysics scenario (left) and for the fixed astrophysics scenario (right), where the astrophysical parameters are fixed to their fiducial value (see table~\ref{tab:params}). We superimpose the forecasted sensitivity from HERA (light blue), Planck (green), and CMB Mix (orange).  The sensitivities from the combination of  HERA and  Planck (CMB Mix) information are represented by the red (dark blue) dashed line. In addition, the red square  represents the parameter space which is promising for solving the $S_8$ tension, coming from Archidiacono et al. analysis for the Planck + BAO dataset~\cite{Archidiacono:2019wdp}. Finally, let us note that we assume a moderate foreground scenario in HERA forecasts.
     \textbf{Bottom:} Same for the combination $a_{\rm dark} \xi^4$, entering in the interaction rate coefficient.  
     }
         \label{fig:constraints_idmdr}
\end{figure}

In the top left panel of figure~\ref{fig:constraints_idmdr}, we present our Fisher matrix forecasts on the $\xi$ parameter, where the  colored curves represent the upper bounds at 95\% confidence level on this parameter. We carried out forecasts for different fixed values of $a_{\rm dark}$, and varied the set of common parameters shown in table~\ref{tab:params}.  The sensitivity forecasts are presented in table~\ref{tab:nidr_cosmo_errors}.
We refer to appendix~\ref{app:derivative} for a discussion on the numerical derivative computation for the $\xi$ parameter. 
We additionally display in appendix~\ref{app:ethos_material} HERA and HERA + CMB Mix full triangle plots and discuss the degeneracies between the DR energy density and the astrophysical parameters.\\

By first looking at the CMB constraints of figure~\ref{fig:constraints_idmdr}, the bounds we obtain from the Planck fisher (green) and the CMB Mix fisher (yellow)
present in both case a degeneracy between $\xi$ and $a_{\rm dark}$, as expected from  their scaling of the interaction rate (see eq.~\ref{eq:Gamma_DMDR}).
The CMB data allow either a small amount of DR with strong interactions, or the opposite, a significant amount of DR weakly interacting with DM. 
In the limit of low-interaction, we find $\xi <0.5736$ with the Fisher analysis of Planck, which can be converted into $\Delta N_{\rm idr} < 0.4170$. 
This forecasted limit is slightly weaker than the one found in section~\ref{sec:neff_section}  ($\Delta N_{\rm eff} = 0.3536$) because we treat DR as a perfect fluid, rather than a free-streaming species,  which decreases its impact on the CMB power spectrum (see solid and dotted black lines on the right panel of figure~\ref{fig:idmdr_PS_Cl}).  The inclusion of SO (\textit{i.e.}, ``CMB Mix'') improves the bound to $\xi < 0.4609$, corresponding to $\Delta N_{\rm idr} < 0.1739 $.\\

We now turn to the HERA forecasts. The Fisher matrix analysis of HERA is shown in figure~\ref{fig:constraints_idmdr} (light blue contours). One can see that HERA is less sensitive to the presence of DR at small $a_{\rm dark}$ than CMB datasets, showing an upper bound of  $\xi < 0.6634$. 
This is similar to what what we found in section~\ref{sec:neff_section}, since the model studied here is (almost) equivalent to the $N_{\rm eff}\Lambda$CDM model in the limit where $a_{\rm dark} \to 0 \, \rm Mpc^{-1}$. However, note that $\xi < 0.6634$ converts to $N_{\rm idr} < 0.7462$, which is $\sim 10\%$ stronger than the bound on $N_{\rm eff}$ reported in section~\ref{sec:neff_section}. We attribute this difference to the fact that $N_{\rm idr}$ is only allowed to take on positive values and thus required to perform one-sided derivative (see eq.~\ref{eq:der_one_sided} in appendix~\ref{app:derivative}). We have checked that enforcing solely positive values to $\Delta N_{\rm eff}$ and making use of one-sided derivatives provides the same bound for HERA as that obtained for $N_{\rm idr}$, indicating that HERA is not sensitive to the streaming properties of the dark radiation itself.\\

When looking at larger amplitude $a_{\rm dark} \gtrsim 10^{-1} \, \rm Mpc^{-1}$, HERA presents a better sensitivity to $\xi$  than Planck alone. Indeed, when increasing the interaction rate, the suppression of the power spectrum at small scales takes an exponential shape, to which the 21cm signal is very sensitive. HERA will thus be very sensitive to models inducing a suppression of this kind, as they have a repercussion on the HMF. 
For the combination of HERA and Planck Fisher matrix (dark blue dotted line), we forecast an upper bound of $ \xi < 0.5040$ in the limit of small $a_{\rm dark}$, \textit{i.e.}, a $\sim 15\%$ gain in sensitivity over Planck alone, and close to that derived from the CMB Mix combination.
In the regime of low interaction rate amplitude, we do not find a significant improvement in combining HERA with CMB Mix, with a sensitivity largely dominated by CMB data. However, for the combination between the CMB and 21cm observables, the increase in sensitivity becomes (again) very significant for larger $a_{\rm dark}$. For the largest amplitude $a_{\rm dark} = 10^{4} \, \rm Mpc^{-1}$ considered in this analysis, the combination of HERA + Planck  is expected to be  one order of magnitude more sensitive to $\xi$ (equivalently , approximately  four orders of magnitude in term of $N_{\rm idr}$) than Planck alone. Including CMB Mix does not improve the sensitivity over HERA + Planck, suggesting rather that at high interaction rate amplitude, the  upper bounds on $\xi$ are almost fully driven by HERA.\\

These behaviors can be understood by looking at the bottom-left panel of figure~\ref{fig:constraints_idmdr}, where we show the  sensitivities (at 95\% C.L.) of the respective experiments on $a_{\rm dark} \xi^4$, the coefficient entering in the interaction rate $\Gamma_{\rm DM-DR} \propto a_{\rm dark} \xi^4$. Contrary to Planck and CMB Mix, HERA's sensitivity is a roughly constant curve in the interval $10^{-1.5}  \lesssim a_{\rm dark} \lesssim 10^4 \, \rm Mpc^{-1}$, indicating that HERA is enforcing a specific limit to the interaction rate affecting DM perturbations, rather than being sensitive to the presence of DR. For $a_{\rm dark} = 10^4 \, \rm Mpc^{-1}$, HERA surpasses CMB sensitivity to the $a_{\rm dark} \xi^4$ product by almost four order of magnitude. \\

The experiment sensitivities to the studied iDMDR model are summarized in table~\ref{tab:nidr_cosmo_errors}, where we present the upper bounds both on the DR energy density $\xi$ (to which CMB experiments are the most sensitive), and the upper bounds on the interaction rate coefficient $a_{\rm dark}\xi^4$ (to which HERA is the most sensitive).\\

\begin{table}[ht]
\centering
   \scalebox{0.85}{
    \begin{NiceTabular}{|c |c | c|c c|}
    \hline 
       &  Planck   &  CMB Mix  & \multicolumn{2}{c|}{HERA}   \\
     \hline
     \diagbox{Param.}{Astro.} & -   & -  & Fixed &  Free   \\ 
        \hline
        $\xi$    &   \hspace{0.5cm} 0.5736   \hspace{0.5cm}      &  0.4609 &   0.6015 & 0.6634         \\
        $a_{\rm dark}\xi^4$ [Mpc$^{-1}$] & 56.65 & 50.72 &1.47$\times 10^{-3}$& 6.12$\times 10^{-3}$ \\
        \hline
         &   \multicolumn{2}{c}{HERA + Planck}    & \multicolumn{2}{c|}{HERA + CMB mix}  \\
     \hline
     \diagbox{Param.}{Astro.} & \multicolumn{2}{c}{Fixed \hspace{1cm} Free } & Fixed& Free  \\ 
        \hline
        $\xi$    &    \multicolumn{2}{c}{  0.4538  \hspace{1cm} 0.5040}                &   0.4329  & 0.4599  \\
        $a_{\rm dark}\xi^4$ [Mpc$^{-1}$]   & \multicolumn{2}{c}{6.94$\times 10^{-4}$ \hspace{1cm} 4.54$\times 10^{-3}$}& 6.41$\times 10^{-4}$ & 4.53$\times 10^{-3}$\\
        \hline
    \end{NiceTabular}
    }
   \caption{Sensitivities (95 \% C.L. upper bound) on $\xi$ and $a_{\rm dark}\xi^4$ from Planck, CMB Mix, and HERA, as well as from their combinations. The $\xi$ upper bounds are taken a for $a_{\rm dark}=10^{-3} \, \rm Mpc^{-1}$, and the $a_{\rm dark}\xi^4$ upper bounds are taken at $a_{\rm dark}=10^4 \, \rm Mpc^{-1}$, the maximum value considered in our analysis. }
    \label{tab:nidr_cosmo_errors}
\end{table}

We also  provide additional Fisher analyses assuming a fixed fiducial astrophysical model for HERA,  presented on the right panels of figure~\ref{fig:constraints_idmdr} (top panel showing $\xi$ and bottom panel showing $a_{\rm dark} \xi^4$), while the upper bounds on sensitivity on $\xi$ and $a_{\rm dark} \xi^4$ are reported in table~\ref{tab:nidr_cosmo_errors}. This allows to gauge the sensitivity to DM-DR interaction that HERA may lead to, if in the future some additional information (\textit{e.g.}, from HST or JWST) allow to strongly constrain the astrophysical parameters entering the model. This is not intended as realistic, but rather as a pedagogic exercise in order to know the most optimistic sensitivity HERA may reach. 
Interestingly, we find that the overall improvement is only of the order of $\sim 10\%$ on the $\xi$ upper bound, suggesting that a realistic modeling of the astrophysics  does not strongly degrade the sensitivity of HERA.  
We refer to  appendix~\ref{app:ethos_material} for a discussion on degeneracies between DR energy density and astrophysical parameters.
It is conceivable that further complexifying the astrophysical model may relax the bound on $\xi$, but we do not expect it to be more severe than the $\sim 10\%$ level that we estimate here. We also verified that the interaction rates corresponding to the HERA 2$\sigma$ forecasted sensitivities do not lead to scale dependent growth of dark matter. However, this effect could become important  moving to real 21cm data analysis (see ref.~\cite{Flitter:2024eay} for the incorporation of a scale-dependent growth factor in \texttt{21cmFirstCLASS}). \\ 

Finally, in figure.~\ref{fig:constraints_idmdr} we display, in the red hatched rectangle, the iDMDR parameter space resulting from the ETHOS n=0 MCMC analysis of ref.~\cite{Archidiacono:2019wdp} for the Planck + BAO dataset, which allows to reach a lower value of $\sigma_8$ than the $\Lambda$CDM fit to Planck CMB data.  This parameter space stretches up to $\xi \simeq 0.16$ for $a_{\rm dark} \ge 1 \, \rm Mpc^{-1}$, corresponding to $\sigma_8 = 0.68$. As can be seen in figure~\ref{fig:constraints_idmdr}, Planck and CMB Mix are not sensitive enough to probe this parameter space,  but our HERA forecasts indicate that future 21cm observations will be able to test these iDMDR scenarios which could alleviate the $S_8$ tension.

\section{Conclusion}
\label{sec:conclusion}

The 21cm signal from CD and EoR offers promising perspectives on the search and characterization of light relics and dark matter properties, notably with the HERA telescope which is gradually approaching detection of the 21cm power spectrum. In this paper, we carried out a series of Fisher matrix analyses (paying particular attention to the calculation of the numerical derivatives to ensure the robustness of our results) in order to estimate the sensitivity of 21cm observations with the HERA telescope to additional free-streaming relativistic species, modeled with $N_{\rm eff}$ and additional dark radiation species interaction with dark matter.
We perform analyses both for HERA alone, and in combination with CMB observations from Planck and the future Simons Observatory, in order to establish the complementarity of both probes.
Our conclusions are as follows.\\

We studied first the effect of  $\Delta N_{\rm eff}$ on the 21cm global signal and power spectrum, and the sensitivity of HERA to $\Delta N_{\rm eff}$: 
\begin{itemize}
    \item Our findings show that HERA alone has a sensitivity that is about 7.5 times  weaker than that of Planck + SO.

\item However, 21cm observations can provide a significant enhancement when mixed with CMB information. Forecasts associating HERA and Planck observations indicate a potential improvement by $\sim  43\%$ on the constraints from Planck alone, and association of HERA and Planck + SO yields a $\sim 23\%$ on the constraints from Planck + SO. 

\item In addition, we find that reducing the astrophysical uncertainties (here done by simply fixing astrophysical parameters) can improve the sensitivity over that of Planck alone (Planck + SO) by a factor $\sim 3.4$ ($\sim 2.8$ respectively). This is a factor of $\sim 2.3$ improvement over the case where the astrophysical parameters are unknown, illustrating the potential synergy between cosmological and astrophysical probes in testing DM-DR interactions.
    
    \item  Finally, and in agreement with past literature, we find that the inclusion of 21cm observations on top of the CMB analyses allows to provide a direct estimate of $\tau_{\rm reio}$, improving in turn the constraints to $A_s$ by $\sim 2.9$  with respect to Planck + SO alone.\\
   
 \end{itemize}

We then studied a model of DM-DR interaction, using the ETHOS parametrisation, focusing on the $n=0$ case. This scenario leads to a smooth suppression in the small scale matter power spectrum, that may explain the $S_8$ tension between CMB and weak lensing surveys. Interestingly, this suppression can strongly impact the astrophysical history, and alter the 21cm observations. Our analysis consisted in estimating the constraints on the DR energy density $\xi$ for different fixed values of DM-DR interaction strength $a_{\rm dark}$, co-varying all cosmological and astrophysical parameters. We conclude that:
\begin{itemize}
    \item Similarly to the case of $N_{\rm eff}$,  HERA alone does not manage to improve the existing upper bound on $\xi$ in the low $a_{\rm dark}$ limit. Nevertheless, HERA and Planck combination provides a $~\sim 15 \%$  sensitivity improvement (which translates in a $68\%$ improvement in $N_{\rm idr}$) compared to Planck alone. 

    \item Furthermore, for large interaction strengths, HERA can  significantly improve the sensitivity to $a_{\rm dark} \xi^4$ by up to four  orders of magnitude over Planck and SO Fisher forecasts.
    
    \item In addition, we found that neglecting astrophysical uncertainties by fixing the astrophysical parameters only improves the sensitivity to $\xi$ by $\sim 10\%$, suggesting a weak dependence of those forecasts on the astrophysical model.

    \item We also found that HERA observations will be able to probe a region of the iDMDR parameter space allowing low values of the $\sigma_8$ parameter, which appear promising to explain the $S_8$ tension. 
\end{itemize}

The combination of HERA and CMB information is therefore highly complementary, and provides sensitivity to the whole DM-DR interaction space ($\xi$, $a_{\rm dark}$). Indeed, CMB data dominate the constraints associated with the effect of the background cosmology, and are the most relevant for imposing constraints to $N_{\rm eff}$ and $N_{\rm idr}$  when the interaction strength is low. On the other hand,  21cm data dominate the constraints associated with the effects of the interaction on DM perturbations, and thus are able to constrain scenarios including a small amount of DR, but high interaction strength. \\

Our analysis can be further extended in several ways, by studying different models of DM-DR interactions, for instance inducing dark acoustic oscillations in the matter power spectrum, as well as through the use of different analysis methods, such as emulators which considerably improve the computational time of high-dimensional inference (see \textit{e.g.}~\cite{Breitman:2023pcj}). Yet, we anticipate that such studies will only strengthen our conclusion that 21cm observations are of prime importance in our quest to investigate the nature and properties of dark matter.\\

\section*{Acknowledgements}
The authors thank Gaétan Facchinetti and Laura Lopez-Honorez for their enlightening discussions, as well as Ely D. Kovetz for his contribution to the conception of this project.
VP and TS are supported by the European Union’s Horizon 2020 research and innovation program under the Marie Sk{\l}odowska-Curie grant agreement no.~860881-HIDDeN. VP is supported by funding from the European Research Council (ERC) under the European Union’s HORIZON-ERC-2022 (grant agreement no.~101076865). JF is supported by the Zin fellowship awarded by the BGU Kreitmann School. The authors acknowledge the use of computational resources from the  LUPM’s cloud computing infrastructure founded by PHONE project - Occitanie Region, Ocevu labex, and France-Grilles.


\appendix

\section{Numerical derivative methodology}
\label{app:derivative}

\subsection{Details on the Fisher forecast for the general case}

As the 21cm power spectrum is computed numerically, it is not possible to find an analytical expression for its derivatives  $\partial \Delta^2_{21} / \partial \theta$  with respect to the parameters of the model. We adopt an approach analogous to \texttt{21cmfish}~\cite{Mason:2022obt} and \texttt{21cmCAST}~\cite{Facchinetti:2023slb}, which consists of calculating the two-sided numerical differentiation by varying each parameter $\theta$ by a few percent around its  fiducial value $\theta_0$: 
\begin{equation}
   \frac{\partial\Delta^2_{21} (k,z)}{\partial \theta} \Big \rvert_{\theta_0} = \frac{\Delta_{21}^2(k,z|\theta_0 + \delta_{\theta}) - \Delta_{21}^2(k,z|\theta_0 - \delta_{\theta})}{2 \delta_{\theta}}\, .
   \label{eq:der_two_sided}
 \end{equation}
The small variation $\delta_{\theta}$ must meet a requirement: the marginalised error $\sigma_{\theta_i} = \sqrt{F_{\theta_i \theta_i}^{-1}}$ shall be independent of $\delta_{\theta_i}$, implying that we need to choose $\delta_{\theta}$  such that the $\sigma_{\theta_i}$  is stable if we slightly vary $\delta_{\theta_i}$. \\

For the astrophysical parameters, we follow the prescription of ref.~\cite{Mason:2022obt}, and adopt $\delta_{\theta} = 3\% \, \theta_0$, where $\theta_0$ is the fiducial value taken by the parameter $\theta$, with the exception of $L_X^{II}$ and $L_X^{III}$ for which we take $\delta_{L_X^{(II,III)}} = \pm 0.1\% \, L_{X,0}^{(II,III)}$.
We also apply $\delta_\theta = 3\% \theta_0$ to  the cosmological parameters after ensuring that their marginalized errors remain stable around $\delta_\theta$. The only exception is $\Delta N_{\rm eff}$ for which we  empirically determine that $\delta_{\Delta N_{\rm eff}} = 2.6\% N_{\rm eff}$ provides more stable results.

\subsection{Details on the Fisher forecast for $\xi$}
In the second part of this paper, we focus on a DM-DR interaction model thanks to the ETHOS parametrisation, and vary the amount of dark radiation, parameterised  by $\xi$ in our analysis.
As the fiducial value of $\xi$  is $\xi_{0} =0$ and only takes positive value considered, we use one-sided numerical differentiation for the DM decay rate parameter in the context of the dark matter decay model:
\begin{equation}
   \frac{\partial\Delta_{21}^2 (k,z)}{\partial \xi} \Big \rvert_{\xi_0} = \frac{\Delta_{21}^2(k,z|\delta_{\xi}) - \Delta_{21}^2(k,z|\xi_0)}{ \delta_{\xi}}\, ,
   \label{eq:der_one_sided}
 \end{equation}
which however offers less precision and stability. 
For the choice of $\delta_{\xi}$,  we use an iterative process, testing values until we find an interval in which $\sigma_{\xi}$ is stable.  To guide us in this task, we refer to a familiar case, using $N_{\rm idr}$, the number of extra relativistic relics:
\begin{equation}
 N_{\rm idr} = \xi^4 (11/4)^{4/3}\, .
\end{equation}  
In the case where the interaction strength $a_{\rm dark} $ is zero, and we are only in the presence of dark radiation, varying $N_{\rm idr}$ is almost equivalent to varying $N_{\rm eff}$, the effective number of ultra-relativistic species.\footnote{The only difference being that we do a fluid approximation in the $n=0$ ETHOS model.}
We can then refer to the previous prescription,  and find the marginalized error is stable close to $\delta_{N_{\rm idr}} = 3 \% N_{\rm eff}$. \footnote{The difference between $\delta_{\Delta N_{\rm eff}}$ and $\delta_{N_{\rm idr}}$ (for $a_{\rm dark} \rightarrow 0 \, \rm Mpc^{-1}$  ) can be explained by the employed derivative methods (two-sided and one-sided derivatives respectively.}
From this starting point, when increasing $a_{\rm dark}$, we decreased $\delta_{N_{\rm idr}}$ in order to maintain stable marginalized errors (larger $a_{\rm dark}$ means that a smaller amount of DR is needed  to be compatible with the observables). We show in table~\ref{tab:delta_nidr}, our choices for $\delta_{N_{\rm idr}}$. We then convert the sensitivity $\sigma_{N_{\rm idr}}$ into the sensitivity for $\xi$ in order to obatin the results of section~\ref{sec:idmdr_section}.\\

\begin{figure}[ht]
     \centering
     \begin{subfigure}[b]{0.49\textwidth}
         \centering
         \includegraphics[width=\textwidth]{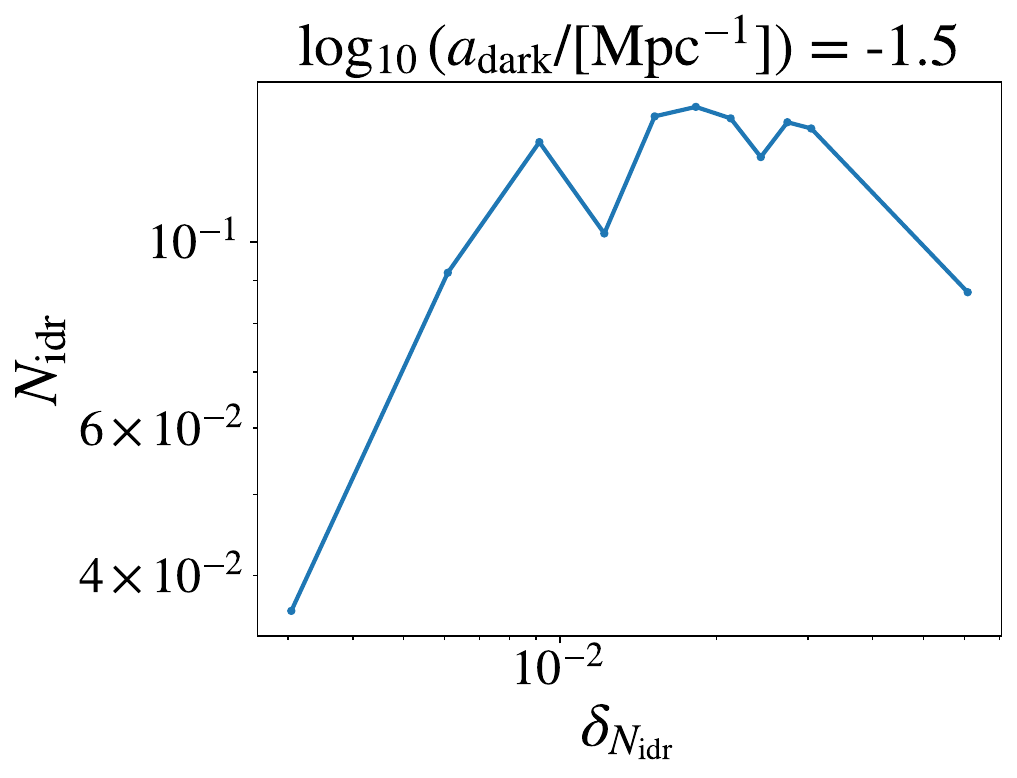}
     \end{subfigure}
     \begin{subfigure}[b]{0.49\textwidth}
         \centering
         \includegraphics[width=\textwidth]{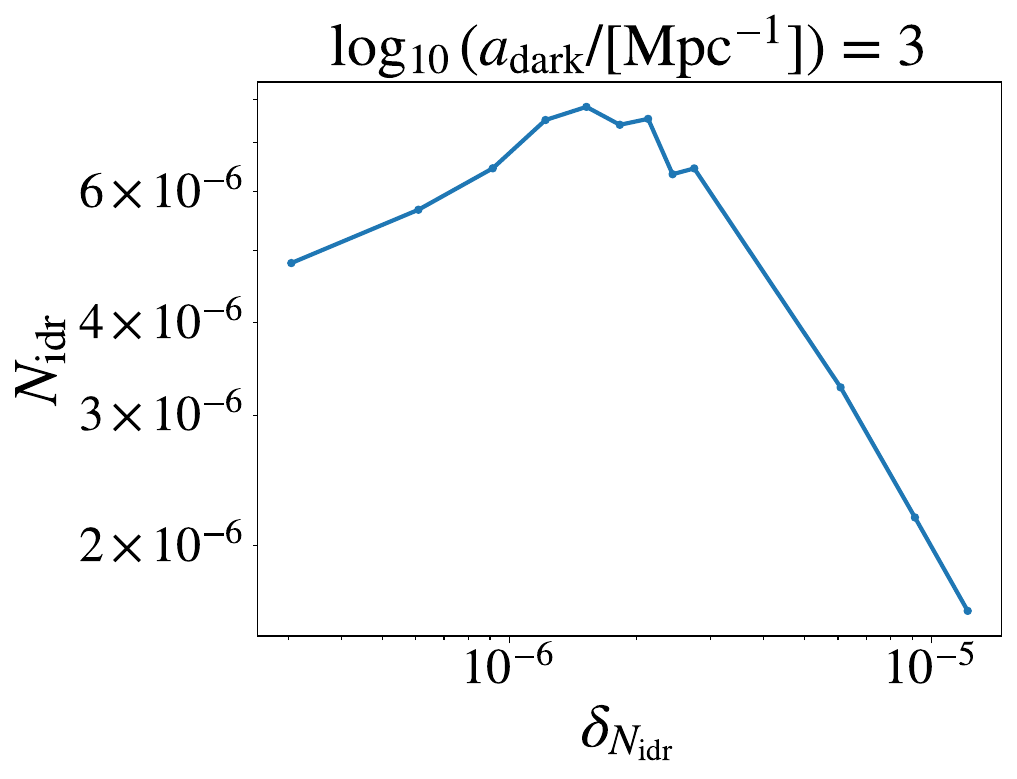}
     \end{subfigure}
        \caption{Sensitivities ($95 \%$ C.L. upper bound) on $N_{\rm idr}$ as a function of $\delta_{N_{\rm idr}}$, the $N_{\rm idr}$ variation chosen for the derivative computation of the 21cm power spectrum, for $ \log_{10} ( a_{\rm dark} /\, \rm [Mpc^{-1}]) = -1.5$ (left) and $3$ (right). The errors are the one 
        from HERA Fisher matrix analysis, varying astrophysical and cosmological parameters and assuming a moderate foreground scenario.}
        \label{fig:deltas}
\end{figure}

As an illustration, we show in figure~\ref{fig:deltas} the $2\sigma$ errors obtained on $N_{\rm idr}$ as a function of $\delta_{N_{\rm idr}}$ from the HERA forecasts, fixing  $a_{\rm dark}$ to $10^{-1.5} \, \rm Mpc^{-1}$ and $10^{3} \, \rm Mpc^{-1}$ while varying all the other cosmological and astrophysical  parameters.
We observe  intervals of stability, where the error on $N_{\rm idr}$ is rather stable  when varying $\delta_{N_{\rm idr}}$. Before these intervals, the model produces too weak modifications on the 21cm power spectrum, compatible with the noise.  After the intervals, the modifications of the power spectrum are too strong,  leading to a non-linear response of the power spectra and invalidating our Fisher approximation. We compute the errors for $\sim 10$ values of $\delta_{N_{\rm idr}}$ for every value of $a_{\rm dark}$.

\begin{table}[ht]
    \centering
    \begin{tabular}{ |c|c c c c|}
    \hline
        $\log_{10} (a_{\rm dark}/  \, [\rm Mpc^{-1}]) $ & -3.0             & -2.5                & -2.0               & -1.5                  \\
         $\delta_{N_{\rm idr}}$ &   9.13$\times 10^{-2}$   &  6.09  $\times 10^{-2}$      &  3.04  $\times 10^{-2}$     &1.83  $\times 10^{-2}$  \\
         \hline
         $\log_{10} (a_{\rm dark} / \, [\rm Mpc^{-1}])$  &   -1.0              & -0.5  &  0.0         &  0.5    \\     
          $\delta_{N_{\rm idr}}$ &  9.13 $\times 10^{-4}$&     6.09 $\times 10^{-3}$    &  2.43 $\times 10^{-3}$ &  1.83 $\times 10^{-4}$    \\
          \hline
          $\log_{10} (a_{\rm dark}/  \, [\rm Mpc^{-1}]) $     & 1.0            &2.0& 3.0& 4.0   \\
         $\delta_{N_{\rm idr}}$ &  9.13 $\times 10^{-5}$ &  1.22 $\times 10^{-5}$ &  1.22 $\times 10^{-6}$ & 2.74 $\times 10^{-7}$\\
         \hline 
    \end{tabular}
    \caption{Values of $\delta_{N_{\rm idr}}$, the variation of $N_{\rm idr}$ to compute the HERA 21cm power spectrum derivative, for each $a_{\rm dark}$ bin included in our analysis.}
    \label{tab:delta_nidr}
\end{table}

\newpage

\section{Supplementary material on $\Delta N_{\rm eff}$}
\label{app:neff_material}
\begin{figure}[H]
    \centering
    \includegraphics[width=1.\linewidth]{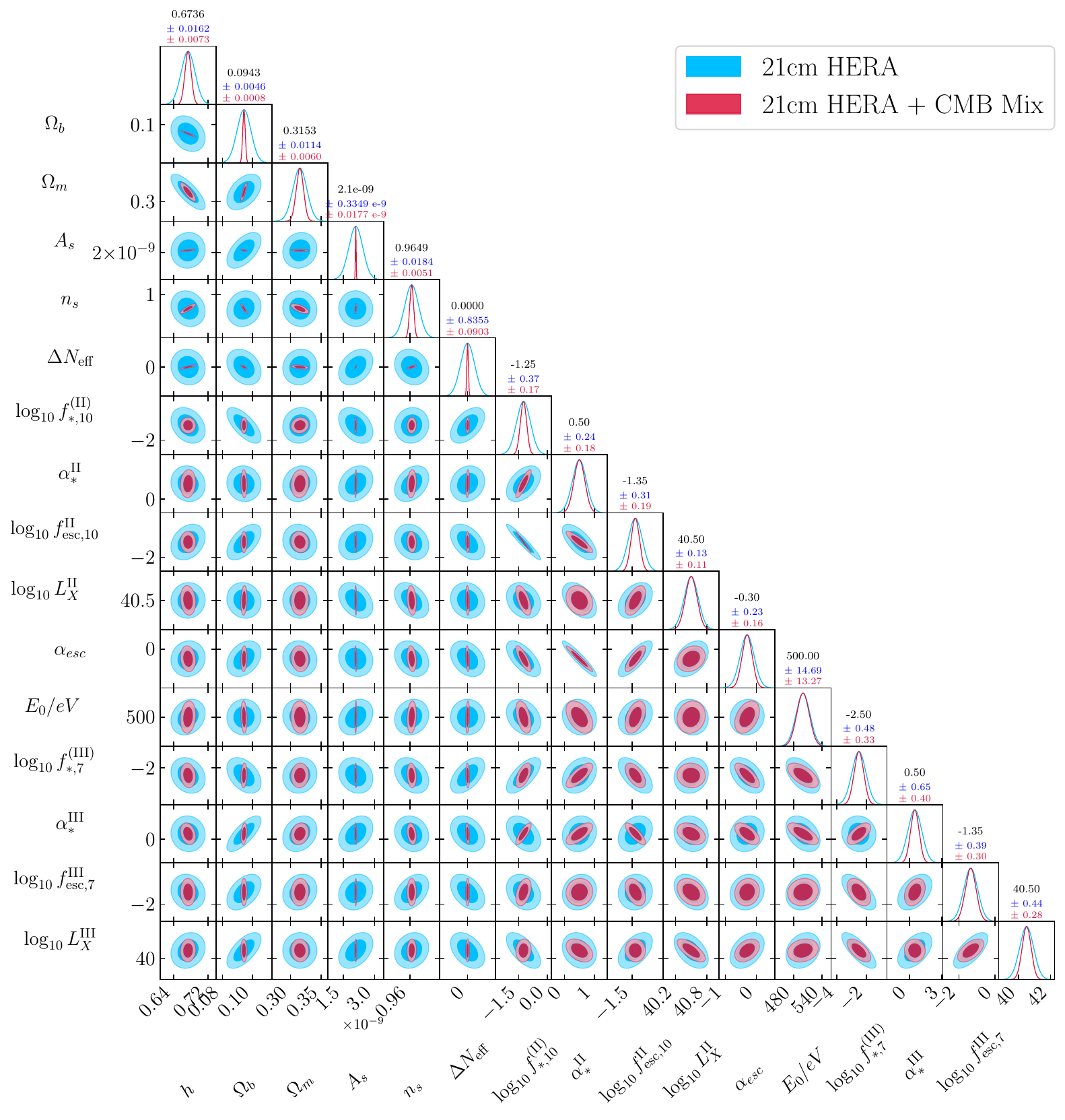}
    \caption{ 1-$\sigma$ and 2-$\sigma$ confidence level forecasts of the $ N_{ \rm eff}\Lambda$CDM model for the HERA and HERA + CMB Mix information.. This figure is similar to figure~\ref{fig:ellipses_neff_21cm_cmbmix}, but we additionally present the astrophysical parameters. Let us recall that the HERA results are computed with a moderate foreground scenario. }
    \label{fig:ellipses_neff_fullastro}
\end{figure}

The upper bounds on $\Delta N_{\rm eff}$ in table~\ref{tab:neff_cosmo_errors}  indicates that a realistic astrophysics modelling degrades HERA sensitivity by $\sim$48\%, compared to a fixed astrophysics scenario, suggesting possible degeneracies between $\Delta N_{\rm eff}$ and astrophysical parameters. Looking at the full triangle plots in figure~\ref{fig:ellipses_neff_fullastro}, the most prominent  $\Delta N_{\rm eff}$ degeneracies can be observed with $f_*$ and $f_{\rm esc}$. Firstly, $f_*$ and $f_{\rm esc}$ are strongly anti correlated with each others [both ACG (II) and MCG parameters (III)], because their product is included in the calculation of the ionization efficiency driving the timing of the astrophysical epochs~\cite{Park:2018ljd, Qin:2020xyh}.
In particular, $\Delta N_{\rm eff}$ presents a correlation with $f_{*,7}^{III}$: increasing $\Delta N_{\rm eff}$ implies a suppression of the low mass halo population, which can be compensated by an increase of the stellar mass in MCGs.  In addition, we observe an anticorrelation between $\Delta N_{\rm eff}$ and the $f_{\rm esc}$ parameters. This behavior is not intuitive, because one could expect that, to compensate a halo population suppression, the fraction of photons escaping the remaining ones to ionize the IGM would increase. However, as pointed out in refs.~\cite{Park:2018ljd,Qin:2020xyh, Munoz:2021psm}, the variation of $f_{\rm esc}$ (II and III), have a very modest impact on the signal, essentially slightly altering the beginning (mostly via $f_{\rm esc,7}^{III}$ associated to MCG) and the ending (mostly via $f_{*,10}^{II}$, associated to ACGs) of the EoR. MCGs are hosted in minihalos, whose population is the most affected by $\Delta N_{\rm eff}$. Hence MCGs parameters are more degenerate with $\Delta N_{\rm eff}$ than the ACGs parameters.\\

A slight anticorrelation is also present between $\Delta N_{\rm eff}$ and $L_X^{III}$. An increased in $L_X^{III}$ impacts the X-ray luminosity, shifting the heating of the IGM to higher redshifts. A positive $\Delta N_{\rm eff}$ delays all the astrophysical epochs, but it also speeds up the transition from one epoch to another.$\Delta N_{\rm eff}$ and $L_X^{III}$ tend to increase the overlap between the Ly$\alpha$ pumping and the X-ray heating, resulting in a reduction of the 21cm absorption feature in the global signal and a suppression of the peaks amplitude in the large scales power spectrum, due to the enhanced negative cross-correlation between the Ly$\alpha$ and IGM temperature fluctuations.\\

All these degeneracies we have indicated here are partial, and a more detailed investigation of the correlation between $\Delta N_{\rm eff}$ and the astrophysical parameters  should be done, by  running analysis for  different models (for instance, without MCGs) and different fiducial values, in order to isolate the  correlations between the parameters. We however leave this question for future work.

\section{Supplementary material on ETHOS}
\label{app:ethos_material}

\begin{figure}[H]
    \centering
    \includegraphics[width=1.\linewidth]{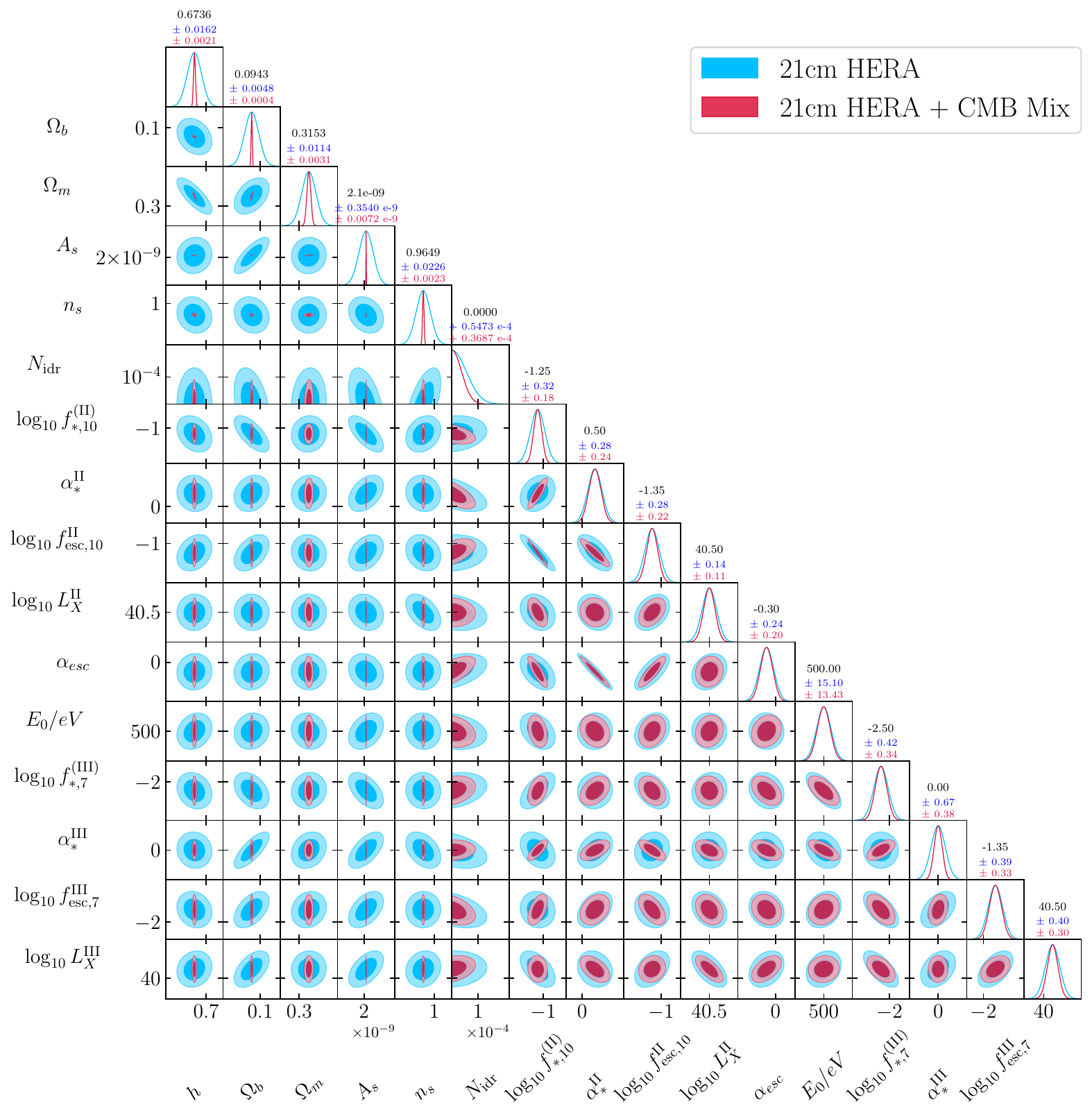}
    \caption{1-$\sigma$ and 2-$\sigma$ confidence level forecasts of the $n=0$ ETHOS model at $  a_{\rm dark} = 10^2 \, \rm Mpc^{-1}$ for the HERA and HERA + CMB Mix information. Let us recall that the HERA results are computed with a moderate foreground scenario.}
    \label{fig:full_dmdr_ellipse}
\end{figure}

We display in figure~\ref{fig:full_dmdr_ellipse} the HERA and HERA + CMB Mix triangle plots for $ a_{\rm dark} =10^2 \, \rm Mpc^{-1}$, for which the sensitivity to DM-DR interactions is dominated by HERA 21cm information. Similar plots can be obtained for different $a_{\rm dark}$. For reasons detailed in appendix~\ref{app:derivative}, the following ellipses feature $N_{\rm idr}$ as the DR energy density parameter instead of $\xi$.  Let us note that the shape of the posteriors are  however not affected by this choice.\\

We can see that $N_{\rm idr}$ presents mild degeneracies with several astrophysical parameters.   
Firstly, $N_{\rm idr}$ is degenerate with 
$f_{*,10}^{III}$: increasing this parameter leads to an increasing stellar mass in MCG, which accelerates the onset of the Cosmic Dawn, and shifts all the 21cm signal to higher redshift, as opposed to the impact of DM-DR interaction, 
implying a correlation between $f_{*,10}^{III}$ and $N_{\rm idr}$.
Moreover, we also observe degeneracies with the $\alpha_{*}^{II}$ and $\alpha_{*}^{III}$ parameters: increasing these parameters results in an enhancement of the stellar-to-halo mass relation for the high mass halos, and giving more power to them in the star formation, implies that the $\alpha_*$ parameters compensate the effect of DM-DR interaction. They are therefore anticorrelated with $N_{\rm idr}$. Finally $N_{ \rm idr}$ is also a slightly anticorrelated with $f_{esc,7}^{III}$, and correlated with $\alpha_{esc}$. This behavior is quite unexpected, because an increase of these parameters tends to bring forward the onset of reionization. However $f_{\rm esc}$ and $\alpha_{esc}$ variations do not leave a strong imprint on the 21cm signal, and only act on the reionization stage, where $N_{\rm idr}$ has less impact.\\
 
Interestingly, the degeneracies between $N_{\rm idr}$ and astrophysical parameters are less marked than in the $\Delta N_{\rm eff}$ analysis (see appendix~\ref{app:neff_material}), which is consistent with the smaller degradation of HERA sensitivity found between a fixed and freely varied astrophysics in the ETHOS analysis.
Further studies are necessary in order to fully characterize all the degeneracies, particularly with $f_{\rm esc}^{III}$ and $\alpha_{\rm esc}$. As already discussed in appendix~\ref{app:neff_material}, isolating the degeneracies with ACG and MCG parameters would be a natural follow-up, that we leave for further study. We would also point out  that by the nature of Fisher analyses, all posteriors are Gaussian, and may miss more complex degeneracies.

\newpage
\bibliographystyle{JHEP}
\bibliography{references}

\providecommand{\href}[2]{#2}\begingroup\raggedright\begin{thebibliography}{100}

\bibitem{Zwicky:1933gu}
F.~Zwicky, \emph{{Die Rotverschiebung von extragalaktischen Nebeln}},
  \href{http://dx.doi.org/10.1007/s10714-008-0707-4}{\emph{Helv. Phys. Acta}
  {\bf 6} (1933) 110--127}.

\bibitem{rubin70}
V.~C. Rubin and J.~We. Kent~Ford, \emph{Rotation of the andromeda nebula from a
  spectroscopic survey of emission regions},
  \href{http://dx.doi.org/10.1086/150317}{\emph{ApJ} {\bf 159} (1970) }.

\bibitem{randall08}
S.~W. {Randall}, M.~{Markevitch}, D.~{Clowe}, A.~H. {Gonzalez} and
  M.~{Brada{\v{c}}}, \emph{{Constraints on the Self-Interaction Cross Section
  of Dark Matter from Numerical Simulations of the Merging Galaxy Cluster 1E
  0657-56}},  \href{http://arxiv.org/abs/0704.0261}{{\tt 0704.0261}}.

\bibitem{Planck:2018vyg}
{\scshape Planck} collaboration, N.~Aghanim et~al., \emph{{Planck 2018 results.
  VI. Cosmological parameters}},
  \href{http://dx.doi.org/10.1051/0004-6361/201833910}{\emph{Astron.
  Astrophys.} {\bf 641} (2020) A6},
  [\href{http://arxiv.org/abs/1807.06209}{{\tt 1807.06209}}].

\bibitem{deBlok:1996jib}
W.~J.~G. de~Blok, S.~S. McGaugh and J.~M. van~der Hulst, \emph{{Hi observations
  of low surface brightness galaxies: probing low density galaxies}},
  \href{http://dx.doi.org/10.1093/mnras/283.1.18}{\emph{Mon. Not. Roy. Astron.
  Soc.} {\bf 283} (1996) 18--54},
  [\href{http://arxiv.org/abs/astro-ph/9605069}{{\tt astro-ph/9605069}}].

\bibitem{Walker:2011zu}
M.~G. Walker and J.~Penarrubia, \emph{{A Method for Measuring (Slopes of) the
  Mass Profiles of Dwarf Spheroidal Galaxies}},
  \href{http://dx.doi.org/10.1088/0004-637X/742/1/20}{\emph{Astrophys. J.} {\bf
  742} (2011) 20}, [\href{http://arxiv.org/abs/1108.2404}{{\tt 1108.2404}}].

\bibitem{Donato:2009ab}
F.~Donato, G.~Gentile, P.~Salucci, C.~F. Martins, M.~I. Wilkinson, G.~Gilmore
  et~al., \emph{{A constant dark matter halo surface density in galaxies}},
  \href{http://dx.doi.org/10.1111/j.1365-2966.2009.15004.x}{\emph{Mon. Not.
  Roy. Astron. Soc.} {\bf 397} (2009) 1169--1176},
  [\href{http://arxiv.org/abs/0904.4054}{{\tt 0904.4054}}].

\bibitem{Oman:2015xda}
K.~A. Oman et~al., \emph{{The unexpected diversity of dwarf galaxy rotation
  curves}}, \href{http://dx.doi.org/10.1093/mnras/stv1504}{\emph{Mon. Not. Roy.
  Astron. Soc.} {\bf 452} (2015) 3650--3665},
  [\href{http://arxiv.org/abs/1504.01437}{{\tt 1504.01437}}].

\bibitem{Boylan-Kolchin:2011qkt}
M.~Boylan-Kolchin, J.~S. Bullock and M.~Kaplinghat, \emph{{Too big to fail? The
  puzzling darkness of massive Milky Way subhaloes}},
  \href{http://dx.doi.org/10.1111/j.1745-3933.2011.01074.x}{\emph{Mon. Not.
  Roy. Astron. Soc.} {\bf 415} (2011) L40},
  [\href{http://arxiv.org/abs/1103.0007}{{\tt 1103.0007}}].

\bibitem{Papastergis:2014aba}
E.~Papastergis, R.~Giovanelli, M.~P. Haynes and F.~Shankar, \emph{{Is there a
  \textquotedblleft{}too big to fail\textquotedblright{} problem in the
  field?}}, \href{http://dx.doi.org/10.1051/0004-6361/201424909}{\emph{Astron.
  Astrophys.} {\bf 574} (2015) A113},
  [\href{http://arxiv.org/abs/1407.4665}{{\tt 1407.4665}}].

\bibitem{Pontzen:2011ty}
A.~Pontzen and F.~Governato, \emph{{How supernova feedback turns dark matter
  cusps into cores}},
  \href{http://dx.doi.org/10.1111/j.1365-2966.2012.20571.x}{\emph{Mon. Not.
  Roy. Astron. Soc.} {\bf 421} (2012) 3464},
  [\href{http://arxiv.org/abs/1106.0499}{{\tt 1106.0499}}].

\bibitem{Garrison-Kimmel:2017zes}
S.~Garrison-Kimmel et~al., \emph{{Not so lumpy after all: modelling the
  depletion of dark matter subhaloes by Milky Way-like galaxies}},
  \href{http://dx.doi.org/10.1093/mnras/stx1710}{\emph{Mon. Not. Roy. Astron.
  Soc.} {\bf 471} (2017) 1709--1727},
  [\href{http://arxiv.org/abs/1701.03792}{{\tt 1701.03792}}].

\bibitem{Heymans:2020gsg}
C.~Heymans et~al., \emph{{KiDS-1000 Cosmology: Multi-probe weak gravitational
  lensing and spectroscopic galaxy clustering constraints}},
  \href{http://dx.doi.org/10.1051/0004-6361/202039063}{\emph{Astron.
  Astrophys.} {\bf 646} (2021) A140},
  [\href{http://arxiv.org/abs/2007.15632}{{\tt 2007.15632}}].

\bibitem{Arico:2023ocu}
G.~Aric\`o, R.~E. Angulo, M.~Zennaro, S.~Contreras, A.~Chen and
  C.~Hern\'andez-Monteagudo, \emph{{DES Y3 cosmic shear down to small scales:
  constraints on cosmology and baryons}},
  \href{http://arxiv.org/abs/2303.05537}{{\tt 2303.05537}}.

\bibitem{Amon:2022azi}
A.~Amon and G.~Efstathiou, \emph{{A non-linear solution to the $S_8$
  tension?}},  \href{http://arxiv.org/abs/2206.11794}{{\tt 2206.11794}}.

\bibitem{Salcido:2024qrt}
J.~Salcido and I.~G. McCarthy, \emph{{Implications of feedback solutions to the
  $S_8$ tension for the baryon fractions of galaxy groups and clusters}},
  \href{http://arxiv.org/abs/2409.05716}{{\tt 2409.05716}}.

\bibitem{Boehm_2001}
C.~Bœhm, P.~Fayet and R.~Schaeffer, \emph{Constraining dark matter candidates
  from structure formation},
  \href{http://dx.doi.org/10.1016/s0370-2693(01)01060-7}{\emph{Physics Letters
  B} {\bf 518} (Oct., 2001) 8–14}.

\bibitem{Boehm_2004}
C.~Bœhm and P.~Fayet, \emph{Scalar dark matter candidates},
  \href{http://dx.doi.org/10.1016/j.nuclphysb.2004.01.015}{\emph{Nuclear
  Physics B} {\bf 683} (Apr., 2004) 219–263}.

\bibitem{Ackerman:2008kmp}
L.~Ackerman, M.~R. Buckley, S.~M. Carroll and M.~Kamionkowski, \emph{{Dark
  Matter and Dark Radiation}},
  \href{http://dx.doi.org/10.1103/PhysRevD.79.023519}{\emph{Phys. Rev. D} {\bf
  79} (2009) 023519}, [\href{http://arxiv.org/abs/0810.5126}{{\tt 0810.5126}}].

\bibitem{Chu:2014lja}
X.~Chu and B.~Dasgupta, \emph{{Dark Radiation Alleviates Problems with Dark
  Matter Halos}},
  \href{http://dx.doi.org/10.1103/PhysRevLett.113.161301}{\emph{Phys. Rev.
  Lett.} {\bf 113} (2014) 161301}, [\href{http://arxiv.org/abs/1404.6127}{{\tt
  1404.6127}}].

\bibitem{Buen-Abad:2015ova}
M.~A. Buen-Abad, G.~Marques-Tavares and M.~Schmaltz, \emph{{Non-Abelian dark
  matter and dark radiation}},
  \href{http://dx.doi.org/10.1103/PhysRevD.92.023531}{\emph{Phys. Rev. D} {\bf
  92} (2015) 023531}, [\href{http://arxiv.org/abs/1505.03542}{{\tt
  1505.03542}}].

\bibitem{Lesgourgues:2015wza}
J.~Lesgourgues, G.~Marques-Tavares and M.~Schmaltz, \emph{{Evidence for dark
  matter interactions in cosmological precision data?}},
  \href{http://dx.doi.org/10.1088/1475-7516/2016/02/037}{\emph{JCAP} {\bf 02}
  (2016) 037}, [\href{http://arxiv.org/abs/1507.04351}{{\tt 1507.04351}}].

\bibitem{Archidiacono:2017slj}
M.~Archidiacono, S.~Bohr, S.~Hannestad, J.~H. J\o{}rgensen and J.~Lesgourgues,
  \emph{{Linear scale bounds on dark matter--dark radiation interactions and
  connection with the small scale crisis of cold dark matter}},
  \href{http://dx.doi.org/10.1088/1475-7516/2017/11/010}{\emph{JCAP} {\bf 11}
  (2017) 010}, [\href{http://arxiv.org/abs/1706.06870}{{\tt 1706.06870}}].

\bibitem{Rubira:2022xhb}
H.~Rubira, A.~Mazoun and M.~Garny, \emph{{Full-shape BOSS constraints on dark
  matter interacting with dark radiation and lifting the S8 tension}},
  \href{http://dx.doi.org/10.1088/1475-7516/2023/01/034}{\emph{JCAP} {\bf 01}
  (2023) 034}, [\href{http://arxiv.org/abs/2209.03974}{{\tt 2209.03974}}].

\bibitem{Mazoun:2023kid}
A.~Mazoun, S.~Bocquet, M.~Garny, J.~J. Mohr, H.~Rubira and S.~M.~L. Vogt,
  \emph{{Probing interacting dark sector models with future weak
  lensing-informed galaxy cluster abundance constraints from SPT-3G and
  CMB-S4}}, \href{http://dx.doi.org/10.1103/PhysRevD.109.063536}{\emph{Phys.
  Rev. D} {\bf 109} (2024) 063536},
  [\href{http://arxiv.org/abs/2312.17622}{{\tt 2312.17622}}].

\bibitem{Mazoun:2024uad}
A.~Mazoun, S.~Bocquet, J.~J. Mohr, M.~Garny, H.~Rubira, M.~Klein et~al.,
  \emph{{Interacting Dark Sector (ETHOS $n=0$): Cosmological Constraints from
  SPT Cluster Abundance with DES and HST Weak Lensing Data}},
  \href{http://arxiv.org/abs/2411.19911}{{\tt 2411.19911}}.

\bibitem{Cyr-Racine:2013fsa}
F.-Y. Cyr-Racine, R.~de~Putter, A.~Raccanelli and K.~Sigurdson,
  \emph{{Constraints on Large-Scale Dark Acoustic Oscillations from
  Cosmology}}, \href{http://dx.doi.org/10.1103/PhysRevD.89.063517}{\emph{Phys.
  Rev. D} {\bf 89} (2014) 063517}, [\href{http://arxiv.org/abs/1310.3278}{{\tt
  1310.3278}}].

\bibitem{Bohr:2020yoe}
S.~Bohr, J.~Zavala, F.-Y. Cyr-Racine, M.~Vogelsberger, T.~Bringmann and
  C.~Pfrommer, \emph{{ETHOS \textendash{} an effective parametrization and
  classification for structure formation: the non-linear regime at z
  \ensuremath{\gtrsim} 5}},
  \href{http://dx.doi.org/10.1093/mnras/staa2579}{\emph{Mon. Not. Roy. Astron.
  Soc.} {\bf 498} (2020) 3403--3419},
  [\href{http://arxiv.org/abs/2006.01842}{{\tt 2006.01842}}].

\bibitem{Archidiacono:2019wdp}
M.~Archidiacono, D.~C. Hooper, R.~Murgia, S.~Bohr, J.~Lesgourgues and M.~Viel,
  \emph{{Constraining Dark Matter-Dark Radiation interactions with CMB, BAO,
  and Lyman-$\alpha$}},
  \href{http://dx.doi.org/10.1088/1475-7516/2019/10/055}{\emph{JCAP} {\bf 10}
  (2019) 055}, [\href{http://arxiv.org/abs/1907.01496}{{\tt 1907.01496}}].

\bibitem{Lopez-Honorez:2016sur}
L.~Lopez-Honorez, O.~Mena, A.~Molin\'e, S.~Palomares-Ruiz and A.~C. Vincent,
  \emph{{The 21 cm signal and the interplay between dark matter annihilations
  and astrophysical processes}},
  \href{http://dx.doi.org/10.1088/1475-7516/2016/08/004}{\emph{JCAP} {\bf 08}
  (2016) 004}, [\href{http://arxiv.org/abs/1603.06795}{{\tt 1603.06795}}].

\bibitem{Lopez-Honorez:2018ipk}
L.~Lopez-Honorez, O.~Mena and P.~Villanueva-Domingo, \emph{{Dark matter
  microphysics and 21 cm observations}},
  \href{http://dx.doi.org/10.1103/PhysRevD.99.023522}{\emph{Phys. Rev. D} {\bf
  99} (2019) 023522}, [\href{http://arxiv.org/abs/1811.02716}{{\tt
  1811.02716}}].

\bibitem{Escudero:2018thh}
M.~Escudero, L.~Lopez-Honorez, O.~Mena, S.~Palomares-Ruiz and
  P.~Villanueva-Domingo, \emph{{A fresh look into the interacting dark matter
  scenario}},
  \href{http://dx.doi.org/10.1088/1475-7516/2018/06/007}{\emph{JCAP} {\bf 06}
  (2018) 007}, [\href{http://arxiv.org/abs/1803.08427}{{\tt 1803.08427}}].

\bibitem{Munoz:2018jwq}
J.~B. Mu\~noz, C.~Dvorkin and A.~Loeb, \emph{{21-cm Fluctuations from Charged
  Dark Matter}},
  \href{http://dx.doi.org/10.1103/PhysRevLett.121.121301}{\emph{Phys. Rev.
  Lett.} {\bf 121} (2018) 121301}, [\href{http://arxiv.org/abs/1804.01092}{{\tt
  1804.01092}}].

\bibitem{Munoz:2018pzp}
J.~B. Mu\~noz and A.~Loeb, \emph{{A small amount of mini-charged dark matter
  could cool the baryons in the early Universe}},
  \href{http://dx.doi.org/10.1038/s41586-018-0151-x}{\emph{Nature} {\bf 557}
  (2018) 684}, [\href{http://arxiv.org/abs/1802.10094}{{\tt 1802.10094}}].

\bibitem{Sun:2023acy}
Y.~Sun, J.~W. Foster, H.~Liu, J.~B. Mu\~noz and T.~R. Slatyer,
  \emph{{Inhomogeneous Energy Injection in the 21-cm Power Spectrum:
  Sensitivity to Dark Matter Decay}},
  \href{http://arxiv.org/abs/2312.11608}{{\tt 2312.11608}}.

\bibitem{Facchinetti:2023slb}
G.~Facchinetti, L.~Lopez-Honorez, Y.~Qin and A.~Mesinger, \emph{{21cm signal
  sensitivity to dark matter decay}},
  \href{http://dx.doi.org/10.1088/1475-7516/2024/01/005}{\emph{JCAP} {\bf 01}
  (2024) 005}, [\href{http://arxiv.org/abs/2308.16656}{{\tt 2308.16656}}].

\bibitem{Qin:2023kkk}
W.~Qin, J.~B. Munoz, H.~Liu and T.~R. Slatyer, \emph{{Birth of the first stars
  amidst decaying and annihilating dark matter}},
  \href{http://dx.doi.org/10.1103/PhysRevD.109.103026}{\emph{Phys. Rev. D} {\bf
  109} (2024) 103026}, [\href{http://arxiv.org/abs/2308.12992}{{\tt
  2308.12992}}].

\bibitem{Flitter:2022pzf}
J.~Flitter and E.~D. Kovetz, \emph{{Closing the window on fuzzy dark matter
  with the 21-cm signal}},
  \href{http://dx.doi.org/10.1103/PhysRevD.106.063504}{\emph{Phys. Rev. D} {\bf
  106} (2022) 063504}, [\href{http://arxiv.org/abs/2207.05083}{{\tt
  2207.05083}}].

\bibitem{Giri:2022nxq}
S.~K. Giri and A.~Schneider, \emph{{Imprints of fermionic and bosonic mixed
  dark matter on the 21-cm signal at cosmic dawn}},
  \href{http://dx.doi.org/10.1103/PhysRevD.105.083011}{\emph{Phys. Rev. D} {\bf
  105} (2022) 083011}, [\href{http://arxiv.org/abs/2201.02210}{{\tt
  2201.02210}}].

\bibitem{wouthuysen52}
S.~A. {Wouthuysen}, \emph{{On the excitation mechanism of the 21-cm
  (radio-frequency) interstellar hydrogen emission line.}}, .

\bibitem{field58}
G.~B. {Field}, \emph{{Excitation of the Hydrogen 21-CM Line}},
  \href{http://dx.doi.org/10.1109/JRPROC.1958.286741}{\emph{Proceedings of the
  IRE} {\bf 46} (Jan., 1958) 240--250}.

\bibitem{DeBoer:2016tnn}
D.~R. DeBoer et~al., \emph{{Hydrogen Epoch of Reionization Array (HERA)}},
  \href{http://dx.doi.org/10.1088/1538-3873/129/974/045001}{\emph{Publ. Astron.
  Soc. Pac.} {\bf 129} (2017) 045001},
  [\href{http://arxiv.org/abs/1606.07473}{{\tt 1606.07473}}].

\bibitem{HERA:2022wmy}
{\scshape HERA} collaboration, Z.~Abdurashidova et~al., \emph{{Improved
  Constraints on the 21 cm EoR Power Spectrum and the X-Ray Heating of the IGM
  with HERA Phase I Observations}},
  \href{http://dx.doi.org/10.3847/1538-4357/acaf50}{\emph{Astrophys. J.} {\bf
  945} (2023) 124}, [\href{http://arxiv.org/abs/2210.04912}{{\tt 2210.04912}}].

\bibitem{Lee:2023uxu}
N.~Lee and S.~C. Hotinli, \emph{{Probing light relics through cosmic dawn}},
  \href{http://dx.doi.org/10.1103/PhysRevD.109.043502}{\emph{Phys. Rev. D} {\bf
  109} (2024) 043502}, [\href{http://arxiv.org/abs/2309.15119}{{\tt
  2309.15119}}].

\bibitem{Flitter:2023mjj}
J.~Flitter and E.~D. Kovetz, \emph{{New tool for 21-cm cosmology. I. Probing
  \ensuremath{\Lambda}CDM and beyond}},
  \href{http://dx.doi.org/10.1103/PhysRevD.109.043512}{\emph{Phys. Rev. D} {\bf
  109} (2024) 043512}, [\href{http://arxiv.org/abs/2309.03942}{{\tt
  2309.03942}}].

\bibitem{Flitter:2023rzv}
J.~Flitter and E.~D. Kovetz, \emph{{New tool for 21-cm cosmology. II.
  Investigating the effect of early linear fluctuations}},
  \href{http://dx.doi.org/10.1103/PhysRevD.109.043513}{\emph{Phys. Rev. D} {\bf
  109} (2024) 043513}, [\href{http://arxiv.org/abs/2309.03948}{{\tt
  2309.03948}}].

\bibitem{Mesinger:2010ne}
A.~Mesinger, S.~Furlanetto and R.~Cen, \emph{{21cmFAST: A Fast, Semi-Numerical
  Simulation of the High-Redshift 21-cm Signal}},
  \href{http://dx.doi.org/10.1111/j.1365-2966.2010.17731.x}{\emph{Mon. Not.
  Roy. Astron. Soc.} {\bf 411} (2011) 955},
  [\href{http://arxiv.org/abs/1003.3878}{{\tt 1003.3878}}].

\bibitem{Murray:2020trn}
S.~G. Murray, B.~Greig, A.~Mesinger, J.~B. Mu\~noz, Y.~Qin, J.~Park et~al.,
  \emph{{21cmFAST v3: A Python-integrated C code for generating 3D realizations
  of the cosmic 21cm signal}},
  \href{http://dx.doi.org/10.21105/joss.02582}{\emph{J. Open Source Softw.}
  {\bf 5} (2020) 2582}, [\href{http://arxiv.org/abs/2010.15121}{{\tt
  2010.15121}}].

\bibitem{Lesgourgues:2011re}
J.~Lesgourgues, \emph{{The Cosmic Linear Anisotropy Solving System (CLASS) I:
  Overview}},  \href{http://arxiv.org/abs/1104.2932}{{\tt 1104.2932}}.

\bibitem{SimonsObservatory:2018koc}
{\scshape Simons Observatory} collaboration, P.~Ade et~al., \emph{{The Simons
  Observatory: Science goals and forecasts}},
  \href{http://dx.doi.org/10.1088/1475-7516/2019/02/056}{\emph{JCAP} {\bf 02}
  (2019) 056}, [\href{http://arxiv.org/abs/1808.07445}{{\tt 1808.07445}}].

\bibitem{Verwohlt:2024efh}
J.~Verwohlt, C.~A. Mason, J.~B. Mu\~noz, F.-Y. Cyr-Racine, M.~Vogelsberger and
  J.~Zavala, \emph{{Separating Dark Acoustic Oscillations from Astrophysics at
  Cosmic Dawn}},  \href{http://arxiv.org/abs/2404.17640}{{\tt 2404.17640}}.

\bibitem{Munoz:2023kkg}
J.~B. Mu\~noz, \emph{{An Effective Model for the Cosmic-Dawn 21-cm Signal}},
  \href{http://arxiv.org/abs/2302.08506}{{\tt 2302.08506}}.

\bibitem{Pritchard:2011xb}
J.~R. Pritchard and A.~Loeb, \emph{{21-cm cosmology}},
  \href{http://dx.doi.org/10.1088/0034-4885/75/8/086901}{\emph{Rept. Prog.
  Phys.} {\bf 75} (2012) 086901}, [\href{http://arxiv.org/abs/1109.6012}{{\tt
  1109.6012}}].

\bibitem{Munoz:2021psm}
J.~B. Mu\~noz, Y.~Qin, A.~Mesinger, S.~G. Murray, B.~Greig and C.~Mason,
  \emph{{The impact of the first galaxies on cosmic dawn and reionization}},
  \href{http://dx.doi.org/10.1093/mnras/stac185}{\emph{Mon. Not. Roy. Astron.
  Soc.} {\bf 511} (2022) 3657--3681},
  [\href{http://arxiv.org/abs/2110.13919}{{\tt 2110.13919}}].

\bibitem{Tegmark:1996yt}
M.~Tegmark, J.~Silk, M.~J. Rees, A.~Blanchard, T.~Abel and F.~Palla, \emph{{How
  small were the first cosmological objects?}},
  \href{http://dx.doi.org/10.1086/303434}{\emph{Astrophys. J.} {\bf 474} (1997)
  1--12}, [\href{http://arxiv.org/abs/astro-ph/9603007}{{\tt
  astro-ph/9603007}}].

\bibitem{Haiman:2006si}
Z.~Haiman and G.~L. Bryan, \emph{{Was Star-Formation Suppressed in
  High-Redshift Minihalos?}},
  \href{http://dx.doi.org/10.1086/506580}{\emph{Astrophys. J.} {\bf 650} (2006)
  7--11}, [\href{http://arxiv.org/abs/astro-ph/0603541}{{\tt
  astro-ph/0603541}}].

\bibitem{Trenti:2009cj}
M.~Trenti and M.~Stiavelli, \emph{{The Formation Rates of Population III Stars
  and Chemical Enrichment of Halos during the Reionization Era}},
  \href{http://dx.doi.org/10.1088/0004-637X/694/2/879}{\emph{Astrophys. J.}
  {\bf 694} (2009) 879--892}, [\href{http://arxiv.org/abs/0901.0711}{{\tt
  0901.0711}}].

\bibitem{Glover:2012gx}
S.~C.~O. Glover, \emph{{The First Stars}},
  \href{http://arxiv.org/abs/1209.2509}{{\tt 1209.2509}}.

\bibitem{Bromm:2013iya}
V.~Bromm, \emph{{Formation of the First Stars}},
  \href{http://dx.doi.org/10.1088/0034-4885/76/11/112901}{\emph{Rept. Prog.
  Phys.} {\bf 76} (2013) 112901}, [\href{http://arxiv.org/abs/1305.5178}{{\tt
  1305.5178}}].

\bibitem{mebane13}
R.~H. Mebane, J.~Mirocha and S.~R. Furlanetto, \emph{{The Persistence of
  Population III Star Formation}},
  \href{http://dx.doi.org/10.1093/mnras/sty1833}{\emph{Monthly Notices of the
  Royal Astronomical Society} {\bf 479} (07, 2018) 4544--4559},
  [\href{http://arxiv.org/abs/https://academic.oup.com/mnras/article-pdf/479/4/4544/25180578/sty1833.pdf}{{\tt
  https://academic.oup.com/mnras/article-pdf/479/4/4544/25180578/sty1833.pdf}}].

\bibitem{Park:2018ljd}
J.~Park, A.~Mesinger, B.~Greig and N.~Gillet, \emph{{Inferring the astrophysics
  of reionization and cosmic dawn from galaxy luminosity functions and the
  21-cm signal}}, \href{http://dx.doi.org/10.1093/mnras/stz032}{\emph{Mon. Not.
  Roy. Astron. Soc.} {\bf 484} (2019) 933--949},
  [\href{http://arxiv.org/abs/1809.08995}{{\tt 1809.08995}}].

\bibitem{Qin:2020pdx}
Y.~Qin, A.~Mesinger, B.~Greig and J.~Park, \emph{{A tale of two sites
  \textendash{} II. Inferring the properties of minihalo-hosted galaxies with
  upcoming 21-cm interferometers}},
  \href{http://dx.doi.org/10.1093/mnras/staa3408}{\emph{Mon. Not. Roy. Astron.
  Soc.} {\bf 501} (2021) 4748--4758},
  [\href{http://arxiv.org/abs/2009.11493}{{\tt 2009.11493}}].

\bibitem{Qin:2020xyh}
Y.~Qin, A.~Mesinger, J.~Park, B.~Greig and J.~B. Mu\~noz, \emph{{A tale of two
  sites \textendash{} I. Inferring the properties of minihalo-hosted galaxies
  from current observations}},
  \href{http://dx.doi.org/10.1093/mnras/staa1131}{\emph{Mon. Not. Roy. Astron.
  Soc.} {\bf 495} (2020) 123--140},
  [\href{http://arxiv.org/abs/2003.04442}{{\tt 2003.04442}}].

\bibitem{frechet43}
M.~Fréchet, \emph{Sur l'extension de certaines evaluations statistiques au cas
  de petits echantillons}, {\emph{Revue de l'Institut International de
  Statistique / Review of the International Statistical Institute} {\bf 11}
  (1943) 182--205}.

\bibitem{darmois45}
G.~Darmois, \emph{Sur les limites de la dispersion de certaines estimations},
  {\emph{Revue de l'Institut International de Statistique / Review of the
  International Statistical Institute} {\bf 13} (1945) 9--15}.

\bibitem{aitken42}
A.~C. Aitken and H.~Silverstone, \emph{Xv.—on the estimation of statistical
  parameters},
  \href{http://dx.doi.org/10.1017/S008045410000618X}{\emph{Proceedings of the
  Royal Society of Edinburgh. Section A. Mathematical and Physical Sciences}
  {\bf 61} (1942) 186–194}.

\bibitem{Mason:2022obt}
C.~A. Mason, J.~B. Mu\~noz, B.~Greig, A.~Mesinger and J.~Park, \emph{{21cmfish:
  Fisher-matrix framework for fast parameter forecasts from the cosmic 21-cm
  signal}},  \href{http://arxiv.org/abs/2212.09797}{{\tt 2212.09797}}.

\bibitem{Pober:2012zz}
J.~C. Pober, A.~R. Parsons, D.~R. DeBoer, P.~McDonald, M.~McQuinn, J.~E.
  Aguirre et~al., \emph{{The Baryon Acoustic Oscillation Broadband and
  Broad-beam Array: Design Overview and Sensitivity Forecasts}},
  \href{http://dx.doi.org/10.1088/0004-6256/145/3/65}{\emph{Astron. J.} {\bf
  145} (2013) 65}, [\href{http://arxiv.org/abs/1210.2413}{{\tt 1210.2413}}].

\bibitem{Pober:2013jna}
J.~C. Pober et~al., \emph{{What Next-Generation 21 cm Power Spectrum
  Measurements Can Teach Us About the Epoch of Reionization}},
  \href{http://dx.doi.org/10.1088/0004-637X/782/2/66}{\emph{Astrophys. J.} {\bf
  782} (2014) 66}, [\href{http://arxiv.org/abs/1310.7031}{{\tt 1310.7031}}].

\bibitem{Lee:2022gz}
N.~Lee, Y.~Ali-Ha\"\i{}moud, N.~Sch\"oneberg and V.~Poulin, \emph{{What It
  Takes to Solve the Hubble Tension through Modifications of Cosmological
  Recombination}},
  \href{http://dx.doi.org/10.1103/PhysRevLett.130.161003}{\emph{Phys. Rev.
  Lett.} {\bf 130} (2023) 161003}, [\href{http://arxiv.org/abs/2212.04494}{{\tt
  2212.04494}}].

\bibitem{Wu:2014hta}
W.~L.~K. Wu, J.~Errard, C.~Dvorkin, C.~L. Kuo, A.~T. Lee, P.~McDonald et~al.,
  \emph{{A Guide to Designing Future Ground-based Cosmic Microwave Background
  Experiments}},
  \href{http://dx.doi.org/10.1088/0004-637X/788/2/138}{\emph{Astrophys. J.}
  {\bf 788} (2014) 138}, [\href{http://arxiv.org/abs/1402.4108}{{\tt
  1402.4108}}].

\bibitem{Munoz:2016owz}
J.~B. Mu\~noz, E.~D. Kovetz, A.~Raccanelli, M.~Kamionkowski and J.~Silk,
  \emph{{Towards a measurement of the spectral runnings}},
  \href{http://dx.doi.org/10.1088/1475-7516/2017/05/032}{\emph{JCAP} {\bf 05}
  (2017) 032}, [\href{http://arxiv.org/abs/1611.05883}{{\tt 1611.05883}}].

\bibitem{Adi:2020qqf}
T.~Adi and E.~D. Kovetz, \emph{{Can conformally coupled modified gravity solve
  the Hubble tension?}},
  \href{http://dx.doi.org/10.1103/PhysRevD.103.023530}{\emph{Phys. Rev. D} {\bf
  103} (2021) 023530}, [\href{http://arxiv.org/abs/2011.13853}{{\tt
  2011.13853}}].

\bibitem{Shmueli:2023box}
G.~Shmueli, D.~Sarkar and E.~D. Kovetz, \emph{{Mitigating the optical depth
  degeneracy in the cosmological measurement of neutrino masses using 21-cm
  observations}},
  \href{http://dx.doi.org/10.1103/PhysRevD.108.083531}{\emph{Phys. Rev. D} {\bf
  108} (2023) 083531}, [\href{http://arxiv.org/abs/2305.07056}{{\tt
  2305.07056}}].

\bibitem{Madhavacheril_2024}
M.~S. Madhavacheril, F.~J. Qu, B.~D. Sherwin, N.~MacCrann, Y.~Li,
  I.~Abril-Cabezas et~al., \emph{The atacama cosmology telescope: Dr6
  gravitational lensing map and cosmological parameters},
  \href{http://dx.doi.org/10.3847/1538-4357/acff5f}{\emph{The Astrophysical
  Journal} {\bf 962} (Feb., 2024) 113}.

\bibitem{Prince:2021fdv}
H.~Prince and J.~Dunkley, \emph{{Compressed Python likelihood for large scale
  temperature and polarization from Planck}},
  \href{http://dx.doi.org/10.1103/PhysRevD.105.023518}{\emph{Phys. Rev. D} {\bf
  105} (2022) 023518}, [\href{http://arxiv.org/abs/2104.05715}{{\tt
  2104.05715}}].

\bibitem{Akita:2020szl}
K.~Akita and M.~Yamaguchi, \emph{{A precision calculation of relic neutrino
  decoupling}},
  \href{http://dx.doi.org/10.1088/1475-7516/2020/08/012}{\emph{JCAP} {\bf 08}
  (2020) 012}, [\href{http://arxiv.org/abs/2005.07047}{{\tt 2005.07047}}].

\bibitem{Froustey:2020mcq}
J.~Froustey, C.~Pitrou and M.~C. Volpe, \emph{{Neutrino decoupling including
  flavour oscillations and primordial nucleosynthesis}},
  \href{http://dx.doi.org/10.1088/1475-7516/2020/12/015}{\emph{JCAP} {\bf 12}
  (2020) 015}, [\href{http://arxiv.org/abs/2008.01074}{{\tt 2008.01074}}].

\bibitem{Bennett:2020zkv}
J.~J. Bennett, G.~Buldgen, P.~F. De~Salas, M.~Drewes, S.~Gariazzo, S.~Pastor
  et~al., \emph{{Towards a precision calculation of $N_{\rm eff}$ in the
  Standard Model II: Neutrino decoupling in the presence of flavour
  oscillations and finite-temperature QED}},
  \href{http://dx.doi.org/10.1088/1475-7516/2021/04/073}{\emph{JCAP} {\bf 04}
  (2021) 073}, [\href{http://arxiv.org/abs/2012.02726}{{\tt 2012.02726}}].

\bibitem{Liu:2015txa}
A.~Liu, J.~R. Pritchard, R.~Allison, A.~R. Parsons, U.~Seljak and B.~D.
  Sherwin, \emph{{Eliminating the optical depth nuisance from the CMB with 21
  cm cosmology}},
  \href{http://dx.doi.org/10.1103/PhysRevD.93.043013}{\emph{Phys. Rev. D} {\bf
  93} (2016) 043013}, [\href{http://arxiv.org/abs/1509.08463}{{\tt
  1509.08463}}].

\bibitem{Qin:2020xrg}
Y.~Qin, V.~Poulin, A.~Mesinger, B.~Greig, S.~Murray and J.~Park,
  \emph{{Reionization inference from the CMB optical depth and E-mode
  polarization power spectra}},
  \href{http://dx.doi.org/10.1093/mnras/staa2797}{\emph{Mon. Not. Roy. Astron.
  Soc.} {\bf 499} (2020) 550--558},
  [\href{http://arxiv.org/abs/2006.16828}{{\tt 2006.16828}}].

\bibitem{Munoz:2024fas}
J.~B. Mu\~noz, J.~Mirocha, J.~Chisholm, S.~R. Furlanetto and C.~Mason,
  \emph{{Reionization after JWST: a photon budget crisis?}},
  \href{http://arxiv.org/abs/2404.07250}{{\tt 2404.07250}}.

\bibitem{HERA:2021noe}
{\scshape HERA} collaboration, Z.~Abdurashidova et~al., \emph{{HERA Phase I
  Limits on the Cosmic 21 cm Signal: Constraints on Astrophysics and Cosmology
  during the Epoch of Reionization}},
  \href{http://dx.doi.org/10.3847/1538-4357/ac2ffc}{\emph{Astrophys. J.} {\bf
  924} (2022) 51}, [\href{http://arxiv.org/abs/2108.07282}{{\tt 2108.07282}}].

\bibitem{HERA:2021bsv}
{\scshape HERA} collaboration, Z.~Abdurashidova et~al., \emph{{First Results
  from HERA Phase I: Upper Limits on the Epoch of Reionization 21 cm Power
  Spectrum}},
  \href{http://dx.doi.org/10.3847/1538-4357/ac1c78}{\emph{Astrophys. J.} {\bf
  925} (2022) 221}, [\href{http://arxiv.org/abs/2108.02263}{{\tt 2108.02263}}].

\bibitem{Lazare:2023jkg}
H.~Lazare, D.~Sarkar and E.~D. Kovetz, \emph{{HERA bound on x-ray luminosity
  when accounting for population III stars}},
  \href{http://dx.doi.org/10.1103/PhysRevD.109.043523}{\emph{Phys. Rev. D} {\bf
  109} (2024) 043523}, [\href{http://arxiv.org/abs/2307.15577}{{\tt
  2307.15577}}].

\bibitem{Blas:2011rf}
D.~Blas, J.~Lesgourgues and T.~Tram, \emph{{The Cosmic Linear Anisotropy
  Solving System (CLASS) II: Approximation schemes}},
  \href{http://dx.doi.org/10.1088/1475-7516/2011/07/034}{\emph{JCAP} {\bf 07}
  (2011) 034}, [\href{http://arxiv.org/abs/1104.2933}{{\tt 1104.2933}}].

\bibitem{Buen-Abad:2017gxg}
M.~A. Buen-Abad, M.~Schmaltz, J.~Lesgourgues and T.~Brinckmann,
  \emph{{Interacting Dark Sector and Precision Cosmology}},
  \href{http://dx.doi.org/10.1088/1475-7516/2018/01/008}{\emph{JCAP} {\bf 01}
  (2018) 008}, [\href{http://arxiv.org/abs/1708.09406}{{\tt 1708.09406}}].

\bibitem{Krall:2017xcw}
R.~Krall, F.-Y. Cyr-Racine and C.~Dvorkin, \emph{{Wandering in the Lyman-alpha
  Forest: A Study of Dark Matter-Dark Radiation Interactions}},
  \href{http://dx.doi.org/10.1088/1475-7516/2017/09/003}{\emph{JCAP} {\bf 09}
  (2017) 003}, [\href{http://arxiv.org/abs/1705.08894}{{\tt 1705.08894}}].

\bibitem{Pan:2018zha}
Z.~Pan, M.~Kaplinghat and L.~Knox, \emph{{Searching for Signatures of Dark
  matter-Dark Radiation Interaction in Observations of Large-scale Structure}},
  \href{http://dx.doi.org/10.1103/PhysRevD.97.103531}{\emph{Phys. Rev. D} {\bf
  97} (2018) 103531}, [\href{http://arxiv.org/abs/1801.07348}{{\tt
  1801.07348}}].

\bibitem{particledatagroup22}
R.~L. {Workman} and {Particle Data Group}, \emph{{Review of Particle Physics}},
  \href{http://dx.doi.org/10.1093/ptep/ptac097}{\emph{Progress of Theoretical
  and Experimental Physics} {\bf 2022} (Aug., 2022) 083C01}.

\bibitem{Munoz:2020mue}
J.~B. Mu\~noz, S.~Bohr, F.-Y. Cyr-Racine, J.~Zavala and M.~Vogelsberger,
  \emph{{ETHOS - an effective theory of structure formation: Impact of dark
  acoustic oscillations on cosmic dawn}},
  \href{http://dx.doi.org/10.1103/PhysRevD.103.043512}{\emph{Phys. Rev. D} {\bf
  103} (2021) 043512}, [\href{http://arxiv.org/abs/2011.05333}{{\tt
  2011.05333}}].

\bibitem{Schneider:2013ria}
A.~Schneider, R.~E. Smith and D.~Reed, \emph{{Halo Mass Function and the Free
  Streaming Scale}}, \href{http://dx.doi.org/10.1093/mnras/stt829}{\emph{Mon.
  Not. Roy. Astron. Soc.} {\bf 433} (2013) 1573},
  [\href{http://arxiv.org/abs/1303.0839}{{\tt 1303.0839}}].

\bibitem{Schneider:2014rda}
A.~Schneider, \emph{{Structure formation with suppressed small-scale
  perturbations}}, \href{http://dx.doi.org/10.1093/mnras/stv1169}{\emph{Mon.
  Not. Roy. Astron. Soc.} {\bf 451} (2015) 3117--3130},
  [\href{http://arxiv.org/abs/1412.2133}{{\tt 1412.2133}}].

\bibitem{Schaeffer:2021qwm}
T.~Schaeffer and A.~Schneider, \emph{{Dark acoustic oscillations: imprints on
  the matter power spectrum and the halo mass function}},
  \href{http://dx.doi.org/10.1093/mnras/stab1116}{\emph{Mon. Not. Roy. Astron.
  Soc.} {\bf 504} (2021) 3773--3786},
  [\href{http://arxiv.org/abs/2101.12229}{{\tt 2101.12229}}].

\bibitem{Benson:2012su}
A.~J. Benson, A.~Farahi, S.~Cole, L.~A. Moustakas, A.~Jenkins, M.~Lovell
  et~al., \emph{{Dark Matter Halo Merger Histories Beyond Cold Dark Matter: I -
  Methods and Application to Warm Dark Matter}},
  \href{http://dx.doi.org/10.1093/mnras/sts159}{\emph{Mon. Not. Roy. Astron.
  Soc.} {\bf 428} (2013) 1774}, [\href{http://arxiv.org/abs/1209.3018}{{\tt
  1209.3018}}].

\bibitem{Leo:2018odn}
M.~Leo, C.~M. Baugh, B.~Li and S.~Pascoli, \emph{{A new smooth-$k$ space filter
  approach to calculate halo abundances}},
  \href{http://dx.doi.org/10.1088/1475-7516/2018/04/010}{\emph{JCAP} {\bf 04}
  (2018) 010}, [\href{http://arxiv.org/abs/1801.02547}{{\tt 1801.02547}}].

\bibitem{Schewtschenko:2014fca}
J.~A. Schewtschenko, R.~J. Wilkinson, C.~M. Baugh, C.~B\oe{}hm and S.~Pascoli,
  \emph{{Dark matter\textendash{}radiation interactions: the impact on dark
  matter haloes}}, \href{http://dx.doi.org/10.1093/mnras/stv431}{\emph{Mon.
  Not. Roy. Astron. Soc.} {\bf 449} (2015) 3587--3596},
  [\href{http://arxiv.org/abs/1412.4905}{{\tt 1412.4905}}].

\bibitem{Bohr:2021bdm}
S.~Bohr, J.~Zavala, F.-Y. Cyr-Racine and M.~Vogelsberger, \emph{{The halo mass
  function and inner structure of ETHOS haloes at high redshift}},
  \href{http://dx.doi.org/10.1093/mnras/stab1758}{\emph{Mon. Not. Roy. Astron.
  Soc.} {\bf 506} (2021) 128--138},
  [\href{http://arxiv.org/abs/2101.08790}{{\tt 2101.08790}}].

\bibitem{sheth01}
R.~K. {Sheth}, H.~J. {Mo} and G.~{Tormen}, \emph{{Ellipsoidal collapse and an
  improved model for the number and spatial distribution of dark matter
  haloes}},  \href{http://arxiv.org/abs/astro-ph/9907024}{{\tt
  astro-ph/9907024}}.

\bibitem{Flitter:2024eay}
J.~Flitter, S.~Libanore and E.~D. Kovetz, \emph{{Does it matter? A more careful
  treatment of density fluctuations in 21-cm simulations}},
  \href{http://arxiv.org/abs/2411.00089}{{\tt 2411.00089}}.

\bibitem{Breitman:2023pcj}
D.~Breitman, A.~Mesinger, S.~G. Murray, D.~Prelogovic, Y.~Qin and R.~Trotta,
  \emph{{21cmemu: an emulator of 21cmfast summary observables}},
  \href{http://dx.doi.org/10.1093/mnras/stad3849}{\emph{Mon. Not. Roy. Astron.
  Soc.} {\bf 527} (2023) 9833--9852},
  [\href{http://arxiv.org/abs/2309.05697}{{\tt 2309.05697}}].

\end{thebibliography}\endgroup
\end{document}